%% file: jeffrieslanl.tex
\documentclass[usegraphicx,useAMS]{mn2e}

\title[A LDB age for NGC 1960]
  {A lithium depletion boundary age of 22 Myr for NGC 1960}
\author[R.~D. Jeffries et al.]
  {R.~D.~Jeffries$^{1}$, Tim Naylor$^{2}$, N~J. Mayne$^{2}$, 
  Cameron P. M. Bell$^2$ and S.~P. Littlefair$^3$\\
$^{1}$  Astrophysics Group, Keele University, Keele, 
      Staffordshire ST5 5BG, UK\\
$^2$ School of Physics, University of Exeter, Exeter EX4 4QL, UK\\
$^3$ Department of Physics and Astronomy, University of Sheffield,
      Sheffield S3 7RH, UK
}
\date{Submitted 29 April 2013}

\pagerange{\pageref{firstpage}--\pageref{lastpage}} \pubyear{2013}

\def\LaTeX{L\kern-.36em\raise.3ex\hbox{a}\kern-.15em
    T\kern-.1667em\lower.7ex\hbox{E}\kern-.125emX}

\begin{document}

\include{latex_macros}

\label{firstpage}

\maketitle

\begin{abstract}
We present a deep Cousins $RI$ photometric survey of the open cluster
NGC~1960, complete to $R_C \simeq 22$, $I_C \simeq 21$, that is used to
select a sample of very low-mass cluster candidates. Gemini
spectroscopy of a subset of these is used to confirm membership
  and locate the age-dependent ``lithium depletion boundary'' (LDB) --
  the luminosity at which lithium remains unburned in its low-mass
  stars.  The LDB implies a cluster age of $22\pm 4$\,Myr
  and is quite insensitive to choice of evolutionary model.  NGC~1960 is
  the youngest cluster for which a LDB age has been estimated and
  possesses a well populated upper main sequence and a rich low-mass
  pre-main sequence.  The LDB age determined here agrees well with
  precise age estimates made for the same cluster based on isochrone
  fits to its high- and low-mass populations. The concordance between
  these three age estimation techniques, that rely on different facets
  of stellar astrophysics at very different masses, is an important
  step towards calibrating the absolute ages of young open clusters and
  lends confidence to ages determined using any one of them.
\end{abstract}


\section{Introduction}

As pre-main sequence (PMS) stars become older, they contract towards
the zero-age main sequence (ZAMS) and their core temperatures rise. If the PMS
star is more massive than about $0.06\,M_{\odot}$, then the core
temperature will eventually become high enough ($T_c \simeq 3\times
10^{6}$\,K) to burn lithium in p, alpha reactions (Bildsten et
al. 1997; Ushomirsky et al. 1998). 
Since the temperature dependence of this reaction is steep,
and as the mixing timescale in fully convective PMS stars is short,
total Li depletion throughout the star should occur very rapidly.  The
lithium depletion boundary (LDB) technique exploits this physics to
determine the ages of young star clusters by establishing the
age-dependent 
luminosity at which Li remains unburned in the atmospheres of their
low-mass members. In principle, LDB ages are both precise and accurate;
observational and theoretical uncertainties typically contribute to errors of
only 10 per cent in the age determination for clusters in the range
10--200\,Myr (Jeffries \& Naylor 2001; Burke, Pinsonneault \& Sills 2004) --
considerably better than other age estimation methods.

Finding the LDB of a cluster entails quantifying the strength of the
Li~{\sc i}~6708\AA\ feature in groups of faint, very low-mass stars,
using spectroscopy with resolving power $R\geq 3000$. This is
observationally challenging; LDB ages have only been estimated for
seven clusters: the Pleiades, ($125\pm 8$\,Myr; Stauffer, Schultz \&
Kirkpatrick 1998), the Alpha Per cluster ($90 \pm 10$\,Myr; Stauffer et
al. 1999), IC 2391 ($50 \pm 5$\,Myr; Barrado y Navascu\'es, Stauffer \&
Jayawardhana 2004), NGC 2547 ($35 \pm 3$\,Myr; Jeffries \&
Oliveira 2005), IC 4665 ($28\pm 5$\,Myr Manzi et al. 2008), Blanco 1
($132 \pm 24$\,Myr; Cargile, James \& Jeffries 2010) and IC 2602
($46^{+6}_{-5}$\,Myr; Dobbie, Lodieu \& Sharp 2010).  However, the few
LDB ages that are known can be used to calibrate other age estimation
methods that are feasible in more distant clusters
or for isolated field stars, but which rely on considerably more
uncertain stellar physics.

For example, LDB ages can be compared with ages determined from the
positions of higher mass stars in the Hertzsprung-Russell (HR) diagram.
This tests, or could possibly calibrate, the required amount of core
convective overshoot -- a phenomenon that has an important effect on
the evolution of high- and intermediate-mass stars 
(e.g. Maeder 1976; Schaller et al. 1992).
Stauffer et al. (1998, 1999) noted that
the LDB ages of the Pleiades and Alpha Per clusters, were significantly
older than their main sequence turn-off ages determined using high-mass
models with no core convective overshoot, but could be brought into
agreement with a moderate amount of overshooting. 

There are alternative age indicators for lower mass stars too.  Fitting
PMS isochrones in the HR diagram as low-mass PMS stars descend their
Hayashi tracks, monitoring the decline of rotation and magnetic
activity with age, and measuring ongoing Li depletion in the
photospheres of G- and K-stars, have all been used as age indicators
(see the review of Soderblom 2010 and references therein).  Their
accuracy and applicability is limited by the uncertain physics of
convection, magnetic fields, mass-loss and spindown in young stars.
LDB determinations for clusters with a range of ages, and where these
other age indicators can also be determined, can help to identify and
calibrate these uncertainties (e.g. Jeffries \& Oliveira 2005; Jeffries
et al. 2009).

In this paper we present a deep photometric catalogue and a LDB age
estimate for NGC 1960, a rich young cluster at $\sim 1$\,kpc, with a
well-populated PMS and many high-mass main sequence
stars. NGC 1960
turns out to be the youngest cluster with a known LDB age and hence a
very valuable addition. However, its distance means that despite its youth, 
the apparent magnitude of the LDB is
as faint as any yet recorded, its detection 
requiring many hours of spectroscopic exposure on the
Gemini-North telescope, suggesting we are approaching the limit of
what can be done with the present generation of 8--10-m telescopes.

In Section 2 we describe previous work on NGC 1960 and review estimates of
the cluster age, distance and reddening.  Section 3 describes a deep,
$R$- and $I$-band photometric survey used to identify candidate
low-mass PMS stars. Section 4 presents Gemini
multi-object spectroscopy of low-mass candidate cluster members,
measuring spectral types and estimating equivalent widths for the
Li~6708\AA\ and H$\alpha$ lines.  Section 5 discusses cluster
membership, locates the LDB and determines the LDB age. In Section 6 we
discuss our result, comparing the ages determined from different
techniques and mass ranges.

\section{NGC 1960: Age, distance and reddening}
\label{age}

\begin{figure}
\includegraphics[width=80mm]{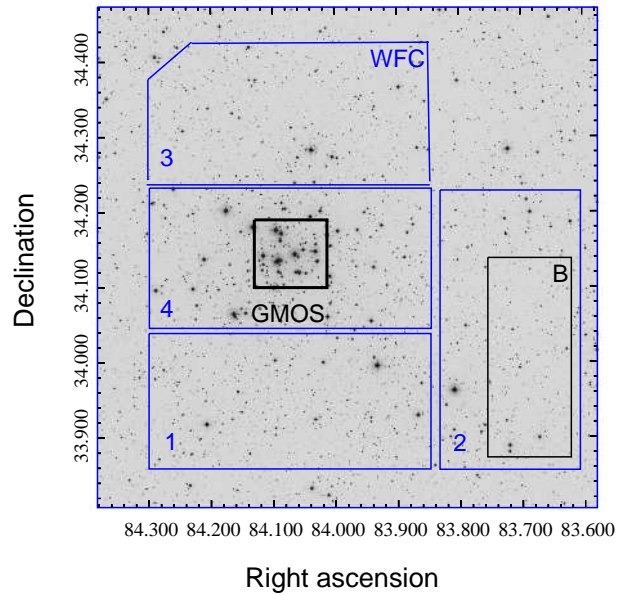}
\caption{A digitized sky survey image around NGC 1960. The four outer
  rectangles (blue in the electronic version) mark the footprint of the $RI$ survey with the Isaac
  Newton Telescope Wide Field Camera (WFC), with CCD numbers
  labelled. No useful data were obtained in the north-east corner of
  CCD 3 because of vignetting.
  The central square (black in the electronic version) inside CCD 4
 marks the Gemini GMOS field in which the
 spectroscopy was taken (see Section~\ref{gmosobs}). The rectangle away
 from the cluster centre marked ``B'' inside CCD 2, is a ``background box'',
 discussed in Section~\ref{member}.
 }
\label{gaiaplot}
\end{figure}

NGC 1960 ($=$ M36) is a rich, northern hemisphere (RA$=$ 05h 36m,
Dec$=+$34d 08m) cluster containing about 15 objects with $V<10$,
corresponding to $M\geq 4M_{\odot}$ at the distance/reddening of the
cluster (see below). The first systematic studies were by
Barkhatova et al. (1985) who used photoelectric photometry 
(from Johnson \& Morgan
1953) and their own photographic $UBV$ photometry to
estimate a reddening $E(B-V)=0.24$, a distance $d=1200$\,pc
and an age of 30\,Myr determined from the main sequence
turn-off. Sanner et al. (2000) present proper motions (to $V=14$) and $BV$ CCD
photometry (to $V=19$), finding a clean cluster main sequence for $V<14$
and determining $E(B-V)=0.25\pm0.02$, $d=1318\pm120$\,pc and an age of
$16^{+10}_{-5}$\,Myr. Sharma et al. (2006) used $UBVRI$ CCD photometry
to determine $E(B-V)=0.22$, $d=1330$\,pc and an age of 25\,Myr. These
authors also examined the radial dependence of surface density in the
cluster, finding a core radius of 3.2 arcminutes and no evidence for
mass segregation. 

Mayne \& Naylor (2008) used the Johnson and
Morgan (1953) photometry and a maximum likelihood fitting technique
to obtain $E(B-V)=0.20\pm0.02$ and $d=1174^{+61}_{-42}$\,pc.
Bell et al. (2013) have adopted a similar maximum likelihood technique
and applied it to both the high- and low-mass populations of NGC~1960,
using updated atmospheres and bolometric corrections and a new method of
applying reddening to stars over a wide range of colours. They used
the $U-B$ versus $B-V$ diagram for the high-mass stars to derive a mean
reddening $E(B-V)=0.20$, with a negligible statistical uncertainty and
no evidence for differential reddening. Applying this reddening 
to the $V$ versus
$B-V$ CMD they obtained a best fit age and intrinsic distance modulus of
 $26.3^{+3.2}_{-5.2}\, \rm{Myr}$ and $10.33^{+0.02}_{-0.05}$ mag
($d=1164^{+11}_{-26}$\,pc). With the distance and reddening fixed at
these values, Bell et al. were then able to fit lower mass cluster
members in the $g$ versus $g-i$ CMD, finding an age of 20\,Myr with 
negligible statistical error, but variations of $\sim 2$\,Myr depending
on which evolutionary models were adopted. The distance
modulus and reddening derived by Bell et al. is used in the rest of this paper.

\section{A CCD Photometric Survey}

\begin{figure}
\includegraphics[width=80mm]{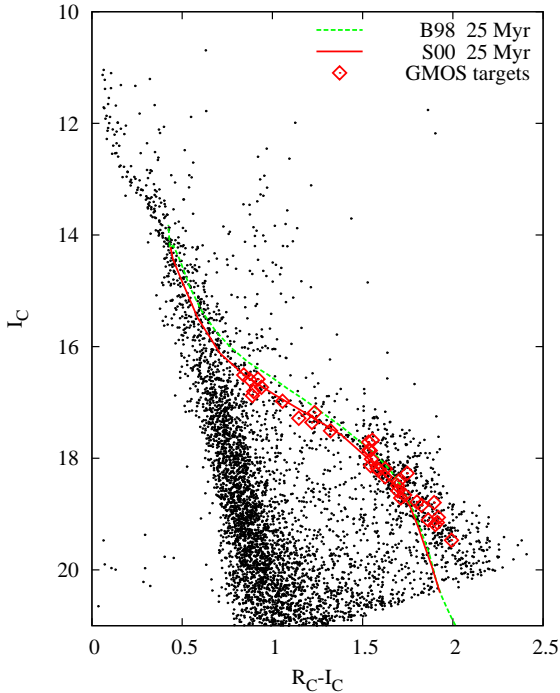}
\caption{A colour-magnitude diagram 
for unflagged objects with uncertainties $<0.1$ mag in $I_C$ and
$R_C-I_C$ seen in CCD 4 (see Fig.~\ref{gaiaplot}).
The dashed and solid lines show theoretical 25\,Myr PMS isochrones (from Baraffe et
al. 1998, with a mixing length of 1.0 pressure scale heights, and from Siess, Dufour \&
Forestini 2000 respectively) at an intrinsic distance modulus
of 10.33, and with a reddening/extinction corresponding to $E(R_C-I_C)=0.143$
(see Section~\ref{age}). The objects for which GMOS spectroscopy were
obtained are indicated.
}
\label{cmd1}
\end{figure}

\begin{table}
\caption{
The range of colours for the photometric standards observed by each CCD detector.
}
\begin{tabular}{ccccccccccccccccc}
\hline
\hline
CCD     & Colour range \cr 
\hline
1       &  0.340 $< R_C-I_C < $ 1.839  \cr
2       &  0.318 $< R_C-I_C < $ 1.750  \cr
3       &  0.342 $< R_C-I_C < $ 2.323  \cr
4       &  0.207 $< R_C-I_C < $ 2.314  \cr
\hline
\hline
\end{tabular}
\label{stand_col}
\end{table}

\begin{table*}
\caption{
The $I_C$ vs $R_C-I_C$ photometric catalogue.
The full table is only available in electronic form, a portion is shown
here to illustrate its content. Columns list the Field and CCD number on which
the star appeared (only one field was observed), a unique identifier on
that CCD, Right Ascension and Declination (J2000.0), the CCD  pixel coordinates
at which the star was found, and then for each of $I_C$ and $R_C-I_C$
there is a magnitude, magnitude uncertainty and a flag (OO for a
``clean'', unflagged detection -- 
a detailed description of the flags is given by Burningham
et al. [2003]).  
}
\begin{tabular}{ccccccccccccc}
\hline
\hline
CCD & ID & RA & Dec & x & y& $I_C$ & $\Delta I_C$ & flag & $R_C-I_C$ &
$\Delta (R_C-I_C)$& flag \\
 &  &   \multicolumn{2}{c}{(J2000.0)} & \multicolumn{2}{c}{(pixels)}  & \multicolumn{2}{c}{(mag)} & &
\multicolumn{2}{c}{(mag)} & \\
\hline
   1.01  &   449&  05 35 44.081& +33 59 44.41&    566.589&   3309.792&      9.431&      0.010&  SS&    0.735&      0.013&  SS\\
   1.02  &   415&  05 35  4.242& +33 54 25.10&   1459.969&   3537.820&      9.649&      0.011&  LS&    0.887&      0.014&  LS\\
   1.04  &   242&  05 36 32.003& +34 10 47.21&    665.336&   1525.915&      9.702&      0.010&  SS&    0.337&      0.013&  SS\\
\hline
\hline
\end{tabular}
\label{ccd_catalogue}
\end{table*}

In order to select faint, low-mass targets for subsequent spectroscopy,
a photometric survey of NGC 1960 was performed using the Wide Field
Camera (WFC)  
at the Isaac Newton Telescope (INT) on La Palma on the
night of the 28th September 2004.  The WFC consists of 4 thinned EEV
2k$\times$4k CCDs (numbered 1--4) covering 0.33 arcsec/pixel on the sky. The
arrangement of the 4 detectors on the sky for our observations of
NGC~1960 is shown in Fig.~\ref{gaiaplot}.  Exposures were obtained in the
Sloan $r$-band (3s, 30s and 3$\times$350s) and Sloan $i$-band (2s, 20s
and 3$\times$200s).  The night was photometric, and so
observations of standard stars from Landolt
(1992) and Stetson (2000) were obtained in the Cousins $RI$ system.  
Table \ref{stand_col} shows the range in
colour of standards observed for each CCD.

The data were de-biassed and flatfielded using master bias and master
twilight sky flat frames. The $i$-band data were defringed using a
library fringe frame.  Photometry was extracted using the optimal
techniques described by Naylor (1998) and Naylor et al. (2002).  The
sum of all three long $i$-band frames was searched to produce a
catalogue of object positions, and then optimal photometry was
performed at these positions in all frames, modelling the background
with a skewed Gaussian distribution (see Burningham et al. 2003).  
By comparing measurements in the long $r$
frames we established that a one percent
magnitude-independent uncertainty should be added 
to measurements from a single frame.
This was included when measurements were combined to yield a single
magnitude for each star in each filter.
The optimal photometry magnitudes were corrected
to that of a large aperture using a spatially dependent aperture
correction (see Naylor et al. 2002).

Standard star photometry was also extracted using optimal photometry
techniques and corrected to a larger aperture in the same way as the
target data. The advantage of this over the more usual method of
performing photometry directly in a large aperture, was that
good signal-to-noise photometry was collected on many more 
faint standards.  The only
disadvantage might be that the standard star magnitudes were originally defined
using a large aperture that included nearby contaminating 
objects. However,
our reduction process flags photometry that is
significantly perturbed by nearby companions and in any case many 
fainter standards (from Stetson 2000) were originally defined 
using PSF-fitting.

The observed standard star instrumental magnitudes were modelled as a
function of colour and airmass to obtain extinction coefficients, zero
points and colour terms.  The airmass range of the standard stars is
small (1.1 to 1.3) and close to the airmass of the target observations
(1.1), and so the extinction was fixed at a single value.  Although a
single linear relationship was sufficient to represent the conversion
from instrumental $i$ to $I_C$ as a function of $R_C-I_C$, we found we
had to use two separate linear relationships to convert instrumental
$r-i$ to $R_C-I_C$, with the break occuring at $R_C-I_C=1.0$ to 1.3
depending on CCD.  A magnitude-independent uncertainty of 1 per cent in
$R_C-I_C$ and 2 per cent in $I_C$
were required to obtain a reduced $\chi^2$ of unity
in our fits.  These values correspond to the combined uncertainty in
the profile correction and correction to the Cousins system.  They
are not included in the uncertainty estimates in the final catalogues,
as they should not be added when comparing stars in a similar region of
the CCD (see Naylor et al. 2002).  The astrometric calibration uses
objects in the 2MASS point source catalogue (Cutri et al. 2003), with a
RMS of 0.1 arcsec for the fit of pixel position as a function of RA and
Dec.

The entire catalogue is presented as Table \ref{ccd_catalogue}, which
is available on-line, or from the Centre de Donn\'ees astronomiques de
Strasbourg (CDS) or from the ``Cluster'' Collaboration's home
page\footnote{www.astro.ex.ac.uk/people/timn/Catalogues/description.html}.
Fig.~\ref{cmd1} shows the $I_C$ vs $R_C-I_C$ colour-magnitude diagram
(CMD) for all unflagged (i.e. clean, star-like, with good photometry)
objects on CCD~4 with colours and magnitudes that have a
signal-to-noise ratio (SNR) greater than 10.  This illustrates a clear
PMS at the position in the CMD appropriate for a $\sim 25$\,Myr
population at a distance of $1164$\,pc and a reddening $E(R_C-I_C)=0.143$
(corresponding to $E(B-V)=0.20$ -- Taylor 1986).
Isochrones are plotted in Fig.~\ref{cmd1} 
from Siess, Dufour \& Forestini (2000, with metallicity of 0.02) 
and Baraffe et al. (1998, with mixing length of 1.0 pressure scale
heights\footnote{Differences due to the adopted mixing length only
  become apparent for masses $>0.5\, M_{\odot}$, or roughly $R_C-I_C<1.3$.}), where the
luminosities and temperatures were 
transformed to the observational plane using a fit to
empirical bolometric corrections from Leggett (1992) and Leggett et
al. (1996) and a colour-$T_{\rm eff}$ relationship that was tuned so
that a 120\,Myr isochrone gave a match to low-mass photometry in the
Pleiades cluster (see Jeffries et al. 2004 for details of this procedure).
The sharp magnitude cut-off in Fig.~\ref{cmd1} is an artefact of the
signal-to-noise threshold placed on the plotted points. We
judge our data to be almost complete down to this cut-off, although the
catalogue detection limit is about 1 magnitude fainter.

\section{Gemini Spectroscopy}

\subsection{Target selection}

\label{gmosobs}

\begin{table*}
\caption{
The identifiers from Table~\ref{ccd_catalogue}, positions and
photometry for the Gemini targets and an integer that indicates in
which slit masks the object was targeted (e.g. 12 indicates that the
object was targeted in masks 1 and 2). The $RI$ photometry is in the
Cousins system and comes from the survey presented here. The gri
photometry are from Bell et al. (2013), except for star 1.04 3081 which
is from a reduction which excludes the deepest i-band image, as a
defect in this image caused the photometry to be flagged.  The gri data
are AB photometric magnitudes calibrated to the natural photometric
system of the INT-WFC (see Bell et al. 2012).}
\begin{tabular}{c@{\hspace*{1mm}}c@{\hspace*{2mm}}c@{\hspace*{1mm}}c@{\hspace*{2mm}}c@{\hspace*{2mm}}c@{\hspace*{1mm}}c@{\hspace*{1mm}}c@{\hspace*{2mm}}c@{\hspace*{1mm}}c@{\hspace*{1mm}}c@{\hspace*{2mm}}c@{\hspace*{1mm}}c@{\hspace*{1mm}}cc}
\hline
\hline
CCD & ID & RA & Dec & $I_C$ & $\Delta I_C$ & $R_C-I_C$ & $\Delta
(R_C-I_C)$ & $i$ & $\Delta i$ & $g-i$ & $\Delta
(g-i)$ & $r-i$ &
$\Delta (r-i)$ & Mask(s) \\
    &    & \multicolumn{2}{c}{(J2000.0)} & \multicolumn{10}{c}{(mag)} & \\
\hline
1.04 &  827 & 5 36 28.151& 34 06 56.30 & 16.583& 0.008 & 0.920& 0.009 &17.086& 0.009&  2.201& 0.015& 0.842& 0.013  &  1   \\  
1.04 &  829 & 5 36 27.929& 34 07 01.14 & 16.591& 0.008 & 0.875& 0.009 &17.063& 0.009&  2.142& 0.015& 0.778& 0.013  &  3   \\ 
1.04 &  876 & 5 36 07.313& 34 09 41.31 & 16.508& 0.008 & 0.841& 0.009 &16.922& 0.009&  1.955& 0.014& 0.717& 0.013  &  1   \\
1.04 &  1018& 5 36 27.761& 34 08 14.77 & 16.806& 0.008 & 0.896& 0.009 &17.289& 0.009&  2.084& 0.015& 0.813& 0.014  &  1   \\
1.04 &  1025& 5 36 25.696& 34 09 52.78 & 16.878& 0.008 & 0.889& 0.009 &17.360& 0.009&  1.972& 0.015& 0.800& 0.014  &  1   \\
1.04 &  1042& 5 36 19.355& 34 10 32.51 & 16.729& 0.012 & 0.939& 0.016 &      &      &       &      &      &        &  3   \\
1.04 &  1056& 5 36 14.361& 34 10 40.92 & 16.979& 0.008 & 1.056& 0.009 &17.496& 0.009&  2.388& 0.017& 0.966& 0.014  &  2   \\
1.04 &  1269& 5 36 24.402& 34 11 32.77 & 17.284& 0.009 & 1.146& 0.010 &17.848& 0.010&  2.578& 0.020& 1.114& 0.015  &  1   \\
1.04 &  1291& 5 36 15.712& 34 10 38.80 & 17.181& 0.009 & 1.234& 0.010 &17.736& 0.010&  2.560& 0.019& 1.158& 0.015  &  1   \\ 
1.04 &  1540& 5 36 13.467& 34 06 56.83 & 17.678& 0.010 & 1.554& 0.012 &18.269& 0.010&  2.865& 0.030& 1.493& 0.017  &  2   \\
1.04 &  1545& 5 36 13.330& 34 11 27.49 & 17.509& 0.009 & 1.322& 0.010 &18.047& 0.010&  2.629& 0.023& 1.231& 0.015  &  3   \\
1.04 &  1833& 5 36 24.869& 34 07 49.36 & 17.357& 0.009 & 1.219& 0.010 &17.919& 0.010&  2.624& 0.022& 1.188& 0.015  &  1   \\
1.04 &  1859& 5 36 17.460& 34 10 50.76 & 17.863& 0.010 & 1.542& 0.013 &18.371& 0.010&  2.875& 0.044& 1.423& 0.020  &  123 \\
1.04 &  1860& 5 36 17.328& 34 11 28.79 & 17.853& 0.010 & 1.535& 0.012 &18.490& 0.020&  2.846& 0.077& 1.459& 0.068  &  2   \\
1.04 &  1871& 5 36 13.418& 34 06 36.46 & 18.015& 0.011 & 1.549& 0.013 &18.602& 0.011&  2.965& 0.041& 1.496& 0.019  &  23  \\
1.04 &  1878& 5 36 10.735& 34 06 32.87 & 17.712& 0.010 & 1.533& 0.012 &18.306& 0.010&  3.029& 0.033& 1.498& 0.017  &  1   \\
1.04 &  2160& 5 36 26.501& 34 11 15.02 & 18.132& 0.011 & 1.544& 0.014 &18.770& 0.011&  2.793& 0.042& 1.480& 0.020  &  3   \\
1.04 &  2171& 5 36 24.976& 34 11 02.18 & 18.191& 0.011 & 1.625& 0.015 &18.999& 0.024&  3.504& 0.297& 1.539& 0.118  &  123 \\ 
1.04 &  2173& 5 36 24.429& 34 09 00.62 & 18.334& 0.012 & 1.625& 0.016 &18.952& 0.011&  3.080& 0.059& 1.621& 0.023  &  123 \\ 
1.04 &  2188& 5 36 19.966& 34 07 46.22 & 18.179& 0.011 & 1.581& 0.015 &18.822& 0.011&  2.899& 0.051& 1.514& 0.022  &  23  \\
1.04 &  2214& 5 36 13.449& 34 06 31.98 & 18.213& 0.012 & 1.587& 0.015 &18.856& 0.011&  3.020& 0.048& 1.568& 0.021  &  23  \\
1.04 &  2249& 5 36 05.655& 34 08 13.25 & 18.270& 0.012 & 1.747& 0.017 &18.928& 0.011&  3.253& 0.069& 1.677& 0.024  &  23  \\
1.04 &  2663& 5 36 16.260& 34 10 12.99 & 18.412& 0.013 & 1.699& 0.018 &19.054& 0.012&  3.076& 0.063& 1.687& 0.025  &  123 \\
1.04 &  2672& 5 36 13.557& 34 11 13.38 & 18.658& 0.014 & 1.733& 0.021 &19.305& 0.022&  3.098& 0.145& 1.758& 0.103  &  12  \\
1.04 &  2696& 5 36 09.583& 34 09 43.17 & 18.584& 0.014 & 1.695& 0.019 &19.240& 0.012&  3.149& 0.080& 1.703& 0.028  &  23  \\
1.04 &  2703& 5 36 07.694& 34 07 10.82 & 18.490& 0.013 & 1.694& 0.018 &19.097& 0.012&  3.222& 0.072& 1.629& 0.024  &  123 \\
1.04 &  3028& 5 36 29.903& 34 09 18.92 & 19.060& 0.019 & 1.919& 0.032 &19.754& 0.014&  3.685& 0.200& 1.921& 0.047  &  123 \\
1.04 &  3073& 5 36 20.090& 34 08 48.03 & 18.785& 0.016 & 1.801& 0.025 &19.460& 0.013&  3.152& 0.095& 1.83 & 0.036  &  123 \\
1.04 &  3080& 5 36 18.218& 34 08 28.28 & 18.795& 0.016 & 1.896& 0.026 &19.442& 0.013&  3.533& 0.143& 1.891& 0.036  &  123 \\
1.04 &  3081& 5 36 18.210& 34 08 03.34 & 18.711& 0.021 & 1.711& 0.077 &19.380& 0.015&  3.210& 0.146& 1.862& 0.051  &  123 \\ 
1.04 &  3150& 5 36 05.662& 34 09 29.78 & 18.843& 0.016 & 1.827& 0.025 &19.532& 0.013&  3.238& 0.113& 1.875& 0.039  &  123 \\
1.04 &  3590& 5 36 18.307& 34 10 24.67 & 19.107& 0.019 & 1.864& 0.033 &19.779& 0.015&  3.786& 0.228& 1.973& 0.053  &  123 \\
1.04 &  3596& 5 36 17.363& 34 10 02.63 & 19.185& 0.020 & 1.903& 0.036 &19.895& 0.015&  3.276& 0.153& 1.756& 0.046  &  123 \\ 
1.04 &  3612& 5 36 13.674& 34 06 45.40 & 19.152& 0.021 & 1.914& 0.037 &19.822& 0.015&  3.536& 0.190& 1.796& 0.046  &  123 \\ 
1.04 &  4165& 5 36 10.731& 34 07 25.67 & 19.469& 0.025 & 1.992& 0.049 &20.213& 0.018&  3.682& 0.275& 2.044& 0.074  &  123 \\
\hline
\hline
\end{tabular}
\label{photomtable}
\end{table*}

The Gemini Multi-Object Spectrograph (GMOS) was used\footnote{Gemini Program
  ID GN-2005B-Q-30} at the Gemini North telescope 
to observe 35 candidate low-mass members of NGC~1960
with $16.5<I_C<19.5$. This corresponds to an approximate 
mass range of $0.15<M/M_{\odot}<0.85$
according to the models of Baraffe et al. (1998). Stars, with unflagged
photometry, were targeted
based on their location in the $I_C, R_C-I_C$ CMD,
following the location of the obvious PMS (see Fig.~\ref{cmd1}).
Targets were
included in three
separate slit mask designs covering the $5.5\times 5.5$ arcmin$^2$ GMOS
field of view,
centered at a single sky position 
 (see Fig.~\ref{gaiaplot}).
The faintest targets were observed through all three masks,
but to cover a larger number of targets, the brighter
candidates were observed through just one or two of the masks.
Table~\ref{photomtable} gives the coordinates and photometry of the
targets and lists which masks they were observed
in.
In addition we list photometry in the $gri_{\rm WFC}$ system from the
photometric survey subsequently performed and detailed in Bell et
al. (2013). This latter survey has slightly poorer precision (and one
target did not have good photometry), but serves as a useful check
on systematic photometric calibration uncertainties. Good 2MASS
photometry was unavailable for most of the faint targets in the sample,
including those around the LDB (see Section~\ref{ldbloc}), so was
not considered.

\subsection{Observations and Data Reduction}

\begin{table}
\caption{Gemini GMOS observation log giving the date and time (UT) at the
  start of each sequence of 3 exposures, the mask number, the
  number of targets in each mask, the exposure times and the average seeing.}
\begin{tabular}{ccccc}
\hline
\hline
Date/Time      & Slit Mask  & $N_{\rm targ}$ & Exposure & Seeing\\
    (UT)       &            &             &     (s)    & (arcsec) \\
\hline
01/11/2005 13:48    & Mask 1  & 26 &  $3\times 1800$& 0.6 \\
06/11/2005 13:05    & Mask 3  & 23 &  $3\times 1800$& 0.6 \\
27/11/2005 08:26    & Mask 2  & 23 &  $3\times 1800$& 0.8 \\
28/11/2005 07:38    & Mask 2  & 23 &  $3\times 1800$& 0.7 \\
02/12/2005 12:34    & Mask 1  & 26 &  $3\times 1800$& 0.6  \\
03/12/2005 12:38    & Mask 1  & 26 &  $3\times 1800$& 0.5 \\  
\hline
\hline
\end{tabular}
\label{slitmask}
\end{table}

\begin{figure*}
\centering
\begin{minipage}[t]{0.47\textwidth}
\includegraphics[width=74mm]{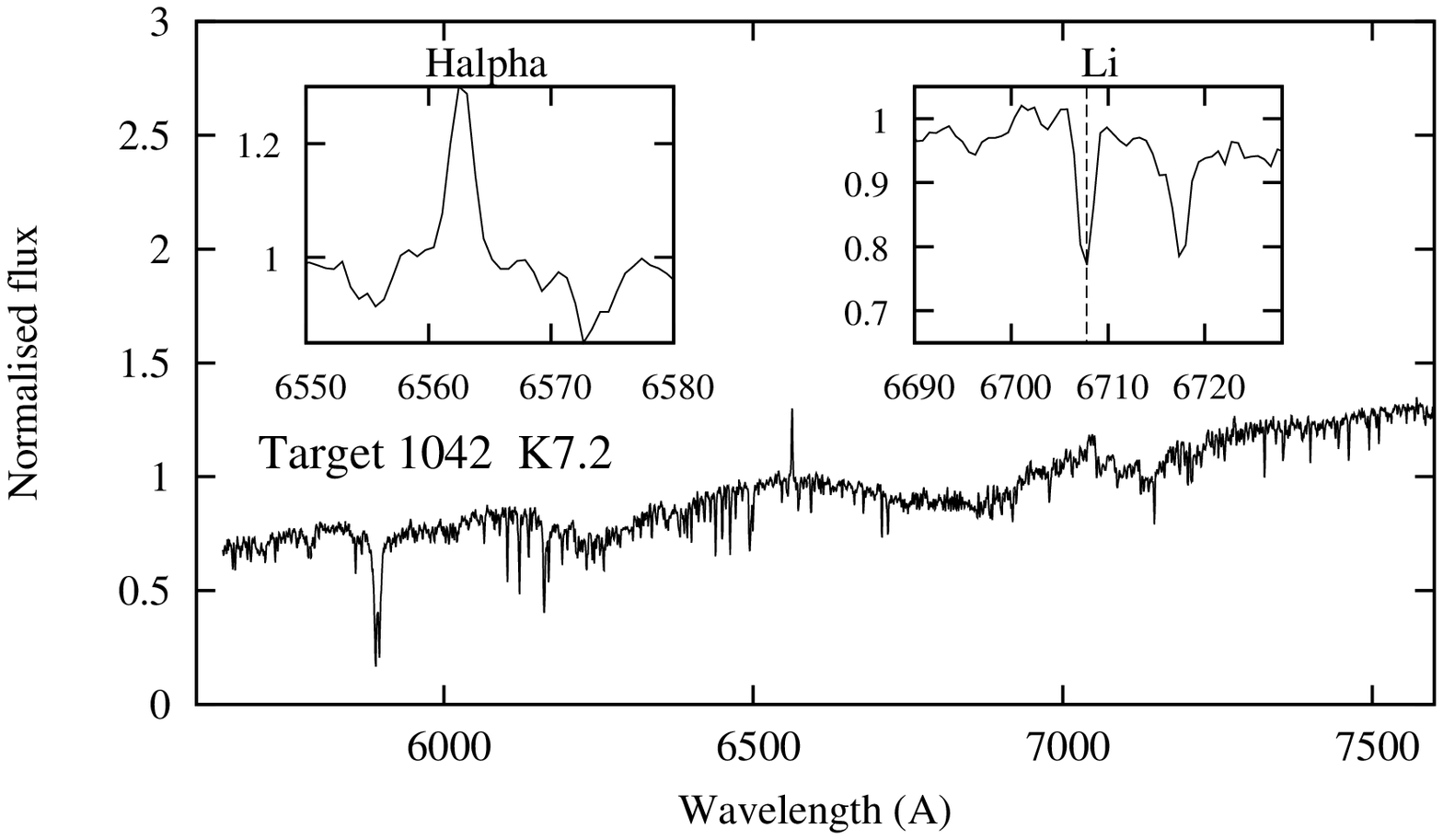}
\includegraphics[width=74mm]{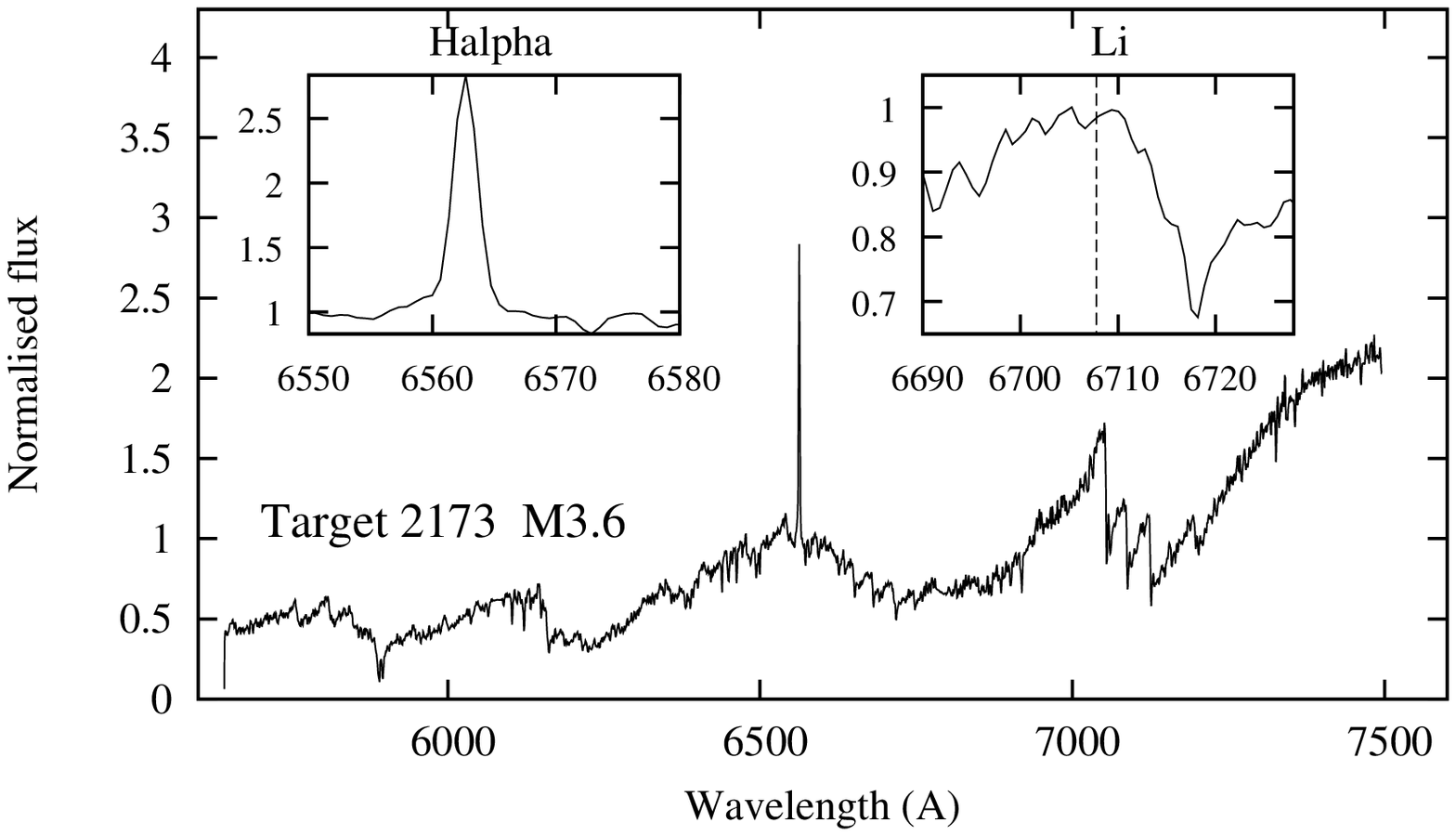}
\includegraphics[width=74mm]{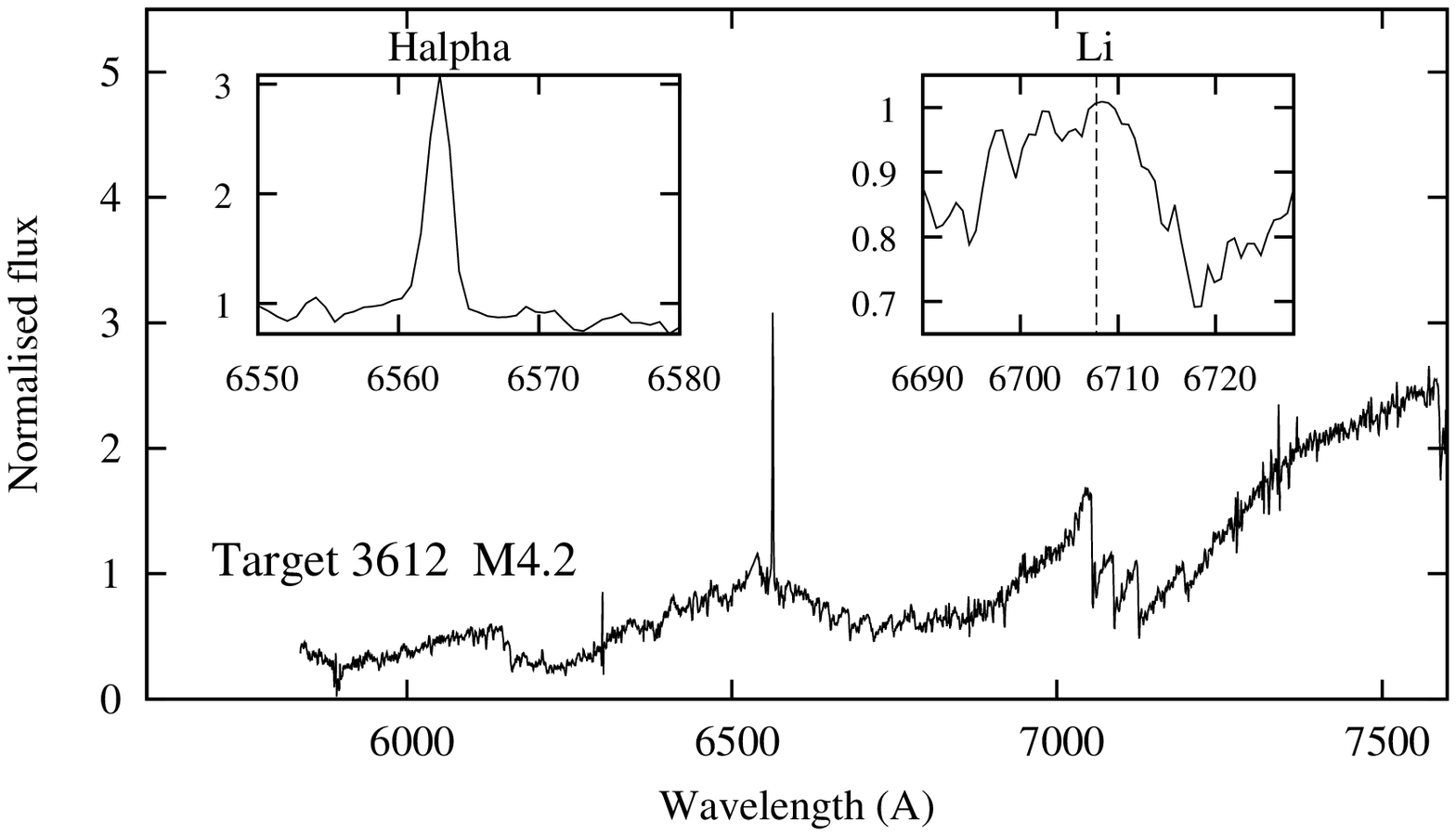}
\includegraphics[width=74mm]{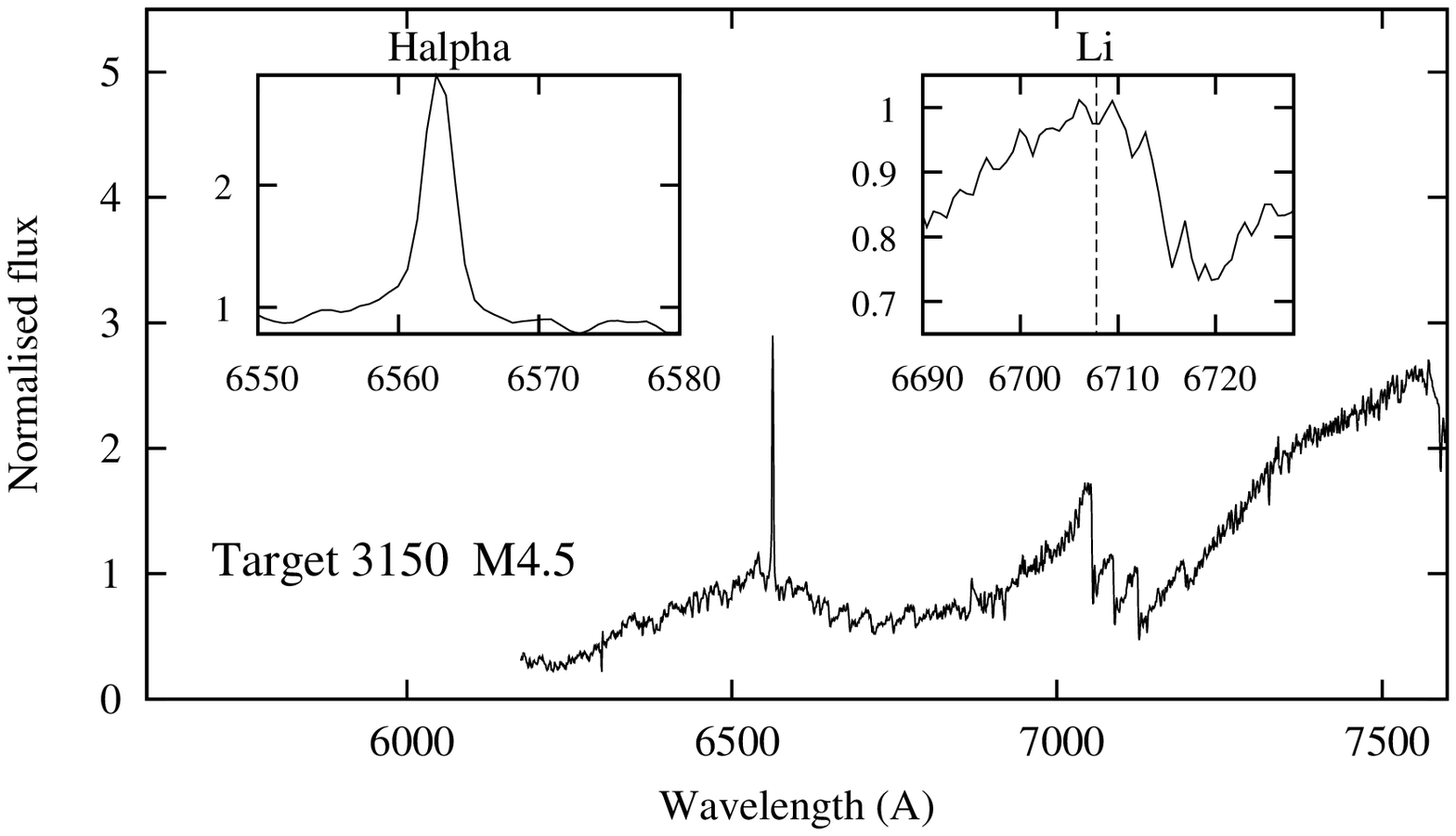}
\includegraphics[width=74mm]{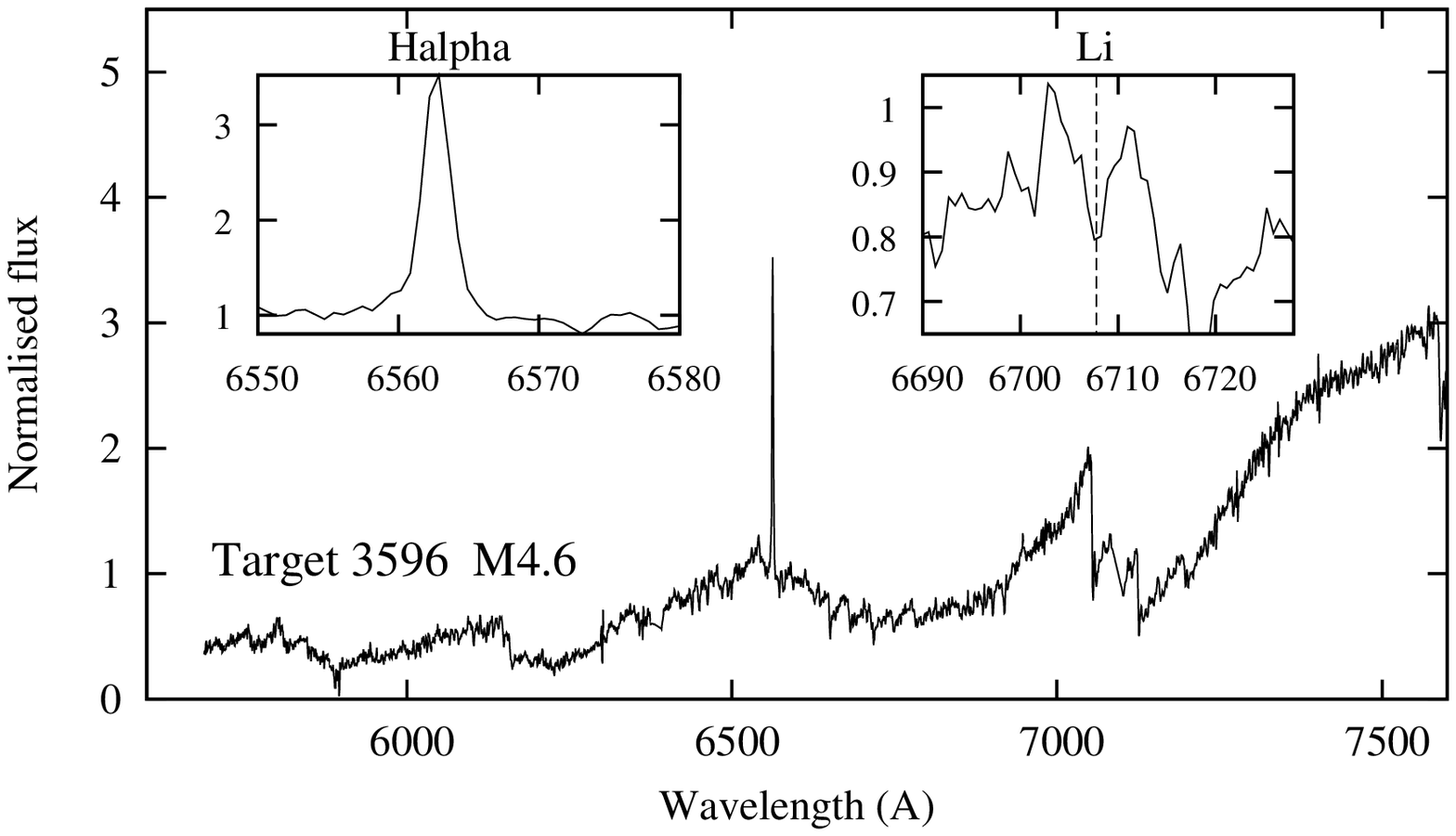}
\end{minipage}
\begin{minipage}[t]{0.47\textwidth}
\includegraphics[width=74mm]{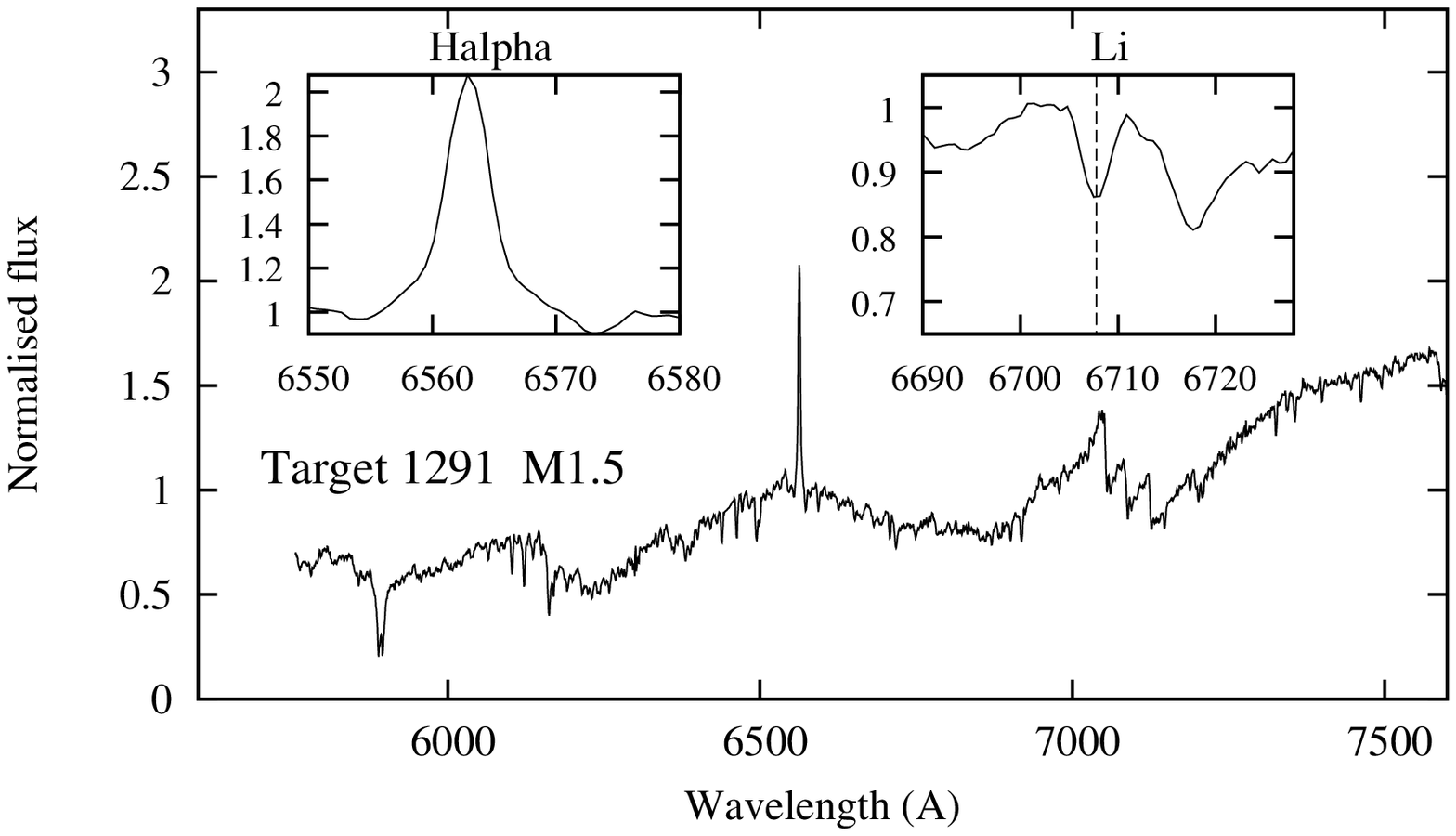}
\includegraphics[width=74mm]{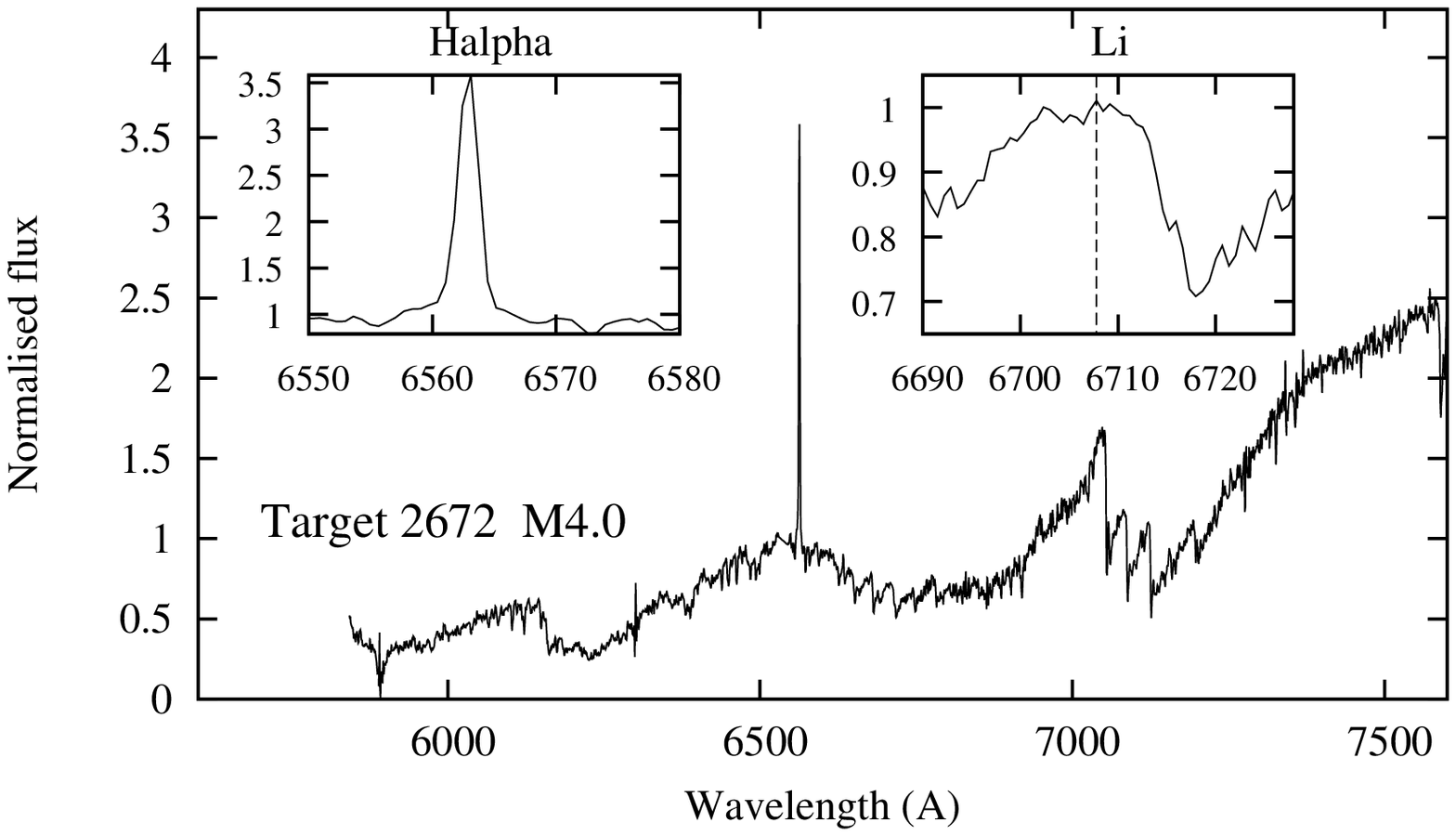}
\includegraphics[width=74mm]{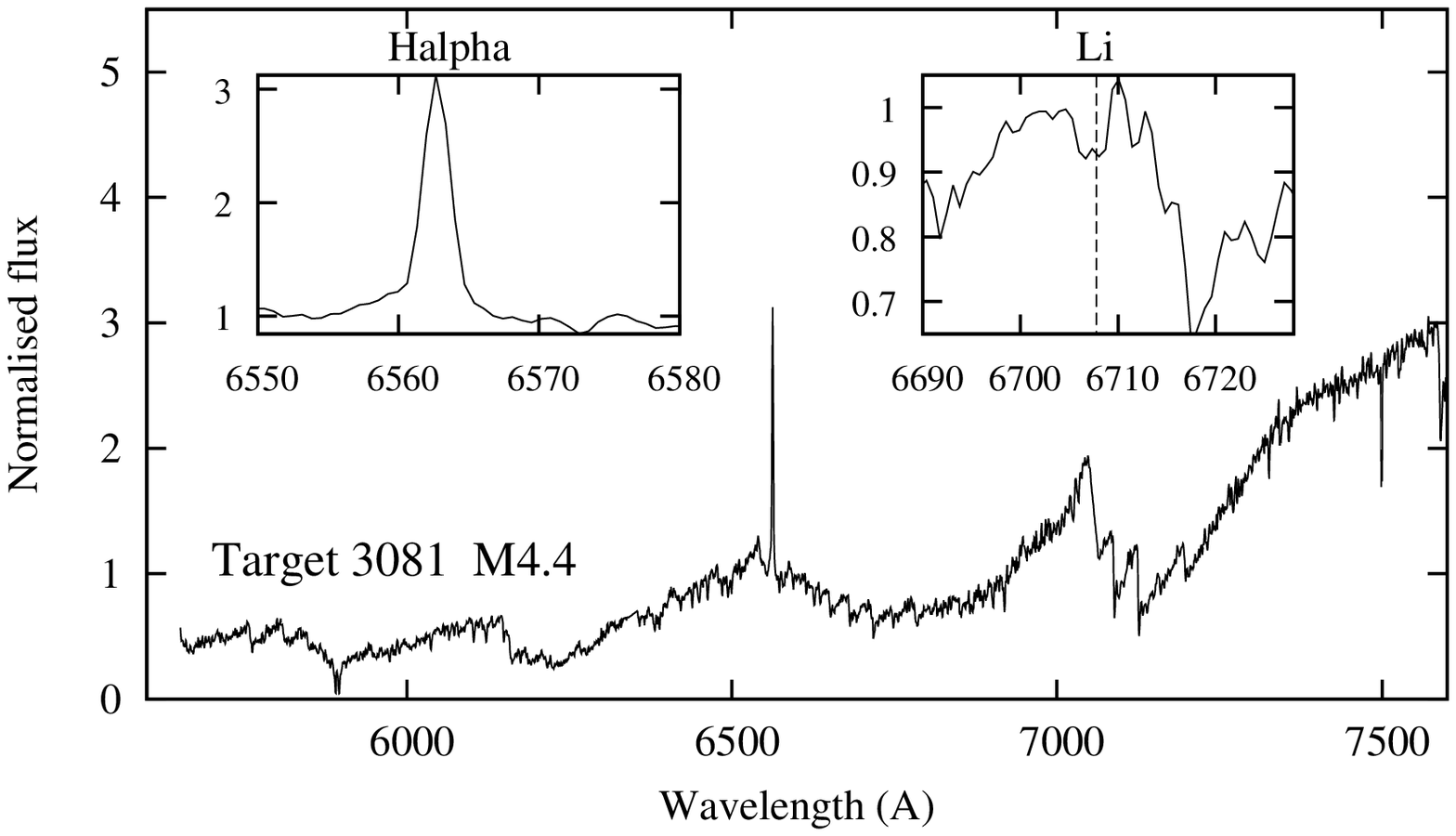}
\includegraphics[width=74mm]{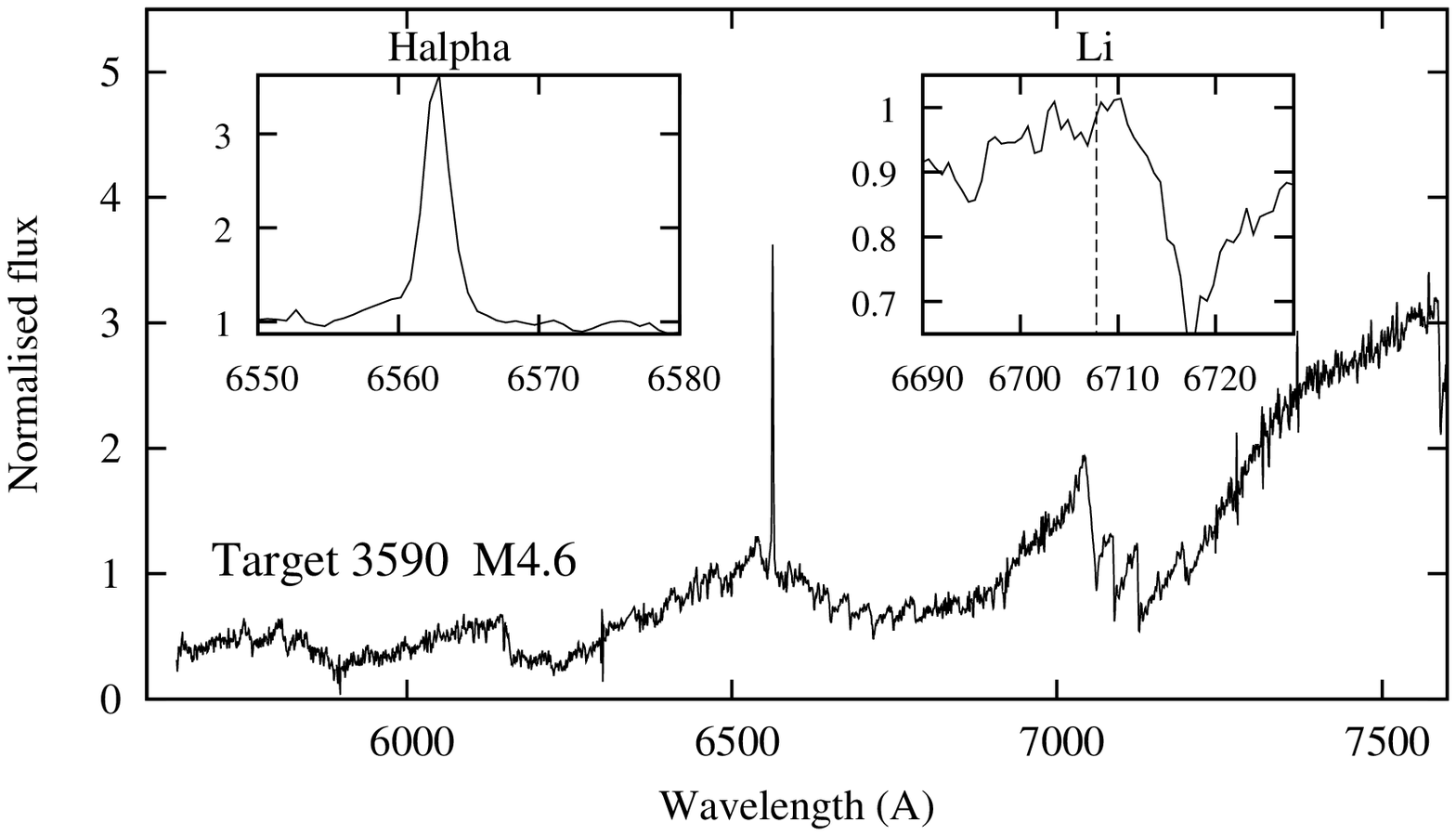}
\includegraphics[width=74mm]{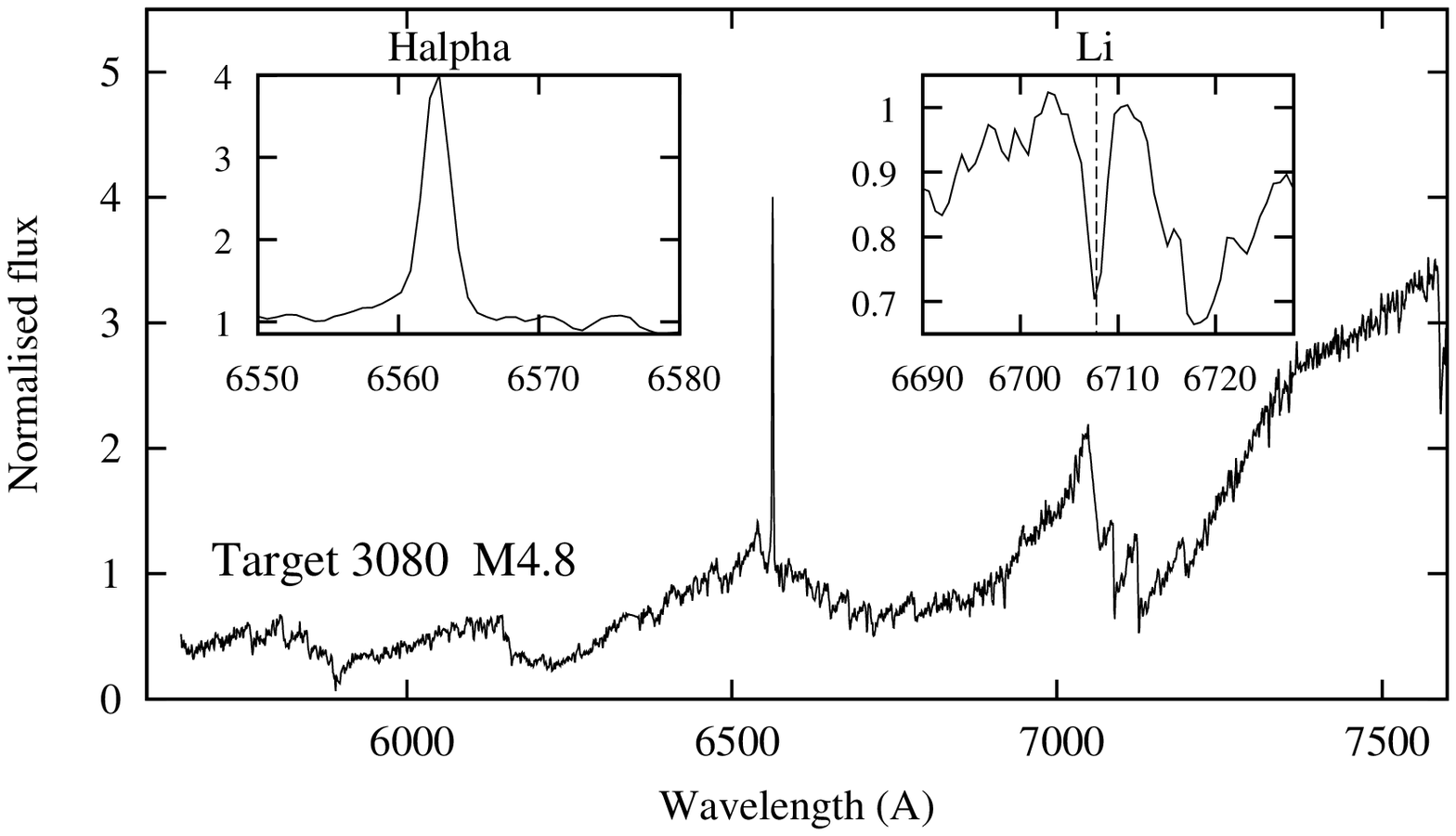}
\end{minipage}
\caption{Example spectra from our target list, covering the full range
  of spectral type and signal-to-noise ratio. Spectra have been subject
  to relative flux calibration, telluric correction and have been
  normalised to a continuum point near H$\alpha$. The inserts on each
  plot show normalised spectra in the regions of the H$\alpha$ and
  Li\,{\sc i}~6708\AA\ lines. Plots are ordered according to spectral
  type (see Section~\ref{spt}) and labelled according to the target ID in Table~\ref{photomtable}.
}
\label{specplot}
\end{figure*}

Each mask setup was observed from one to three times. 
The observations were taken in queue mode
during November and December 2005 (see
Table~\ref{slitmask}). On each occasion that a mask was observed, 
we obtained $3\times1800$\,s
exposures bracketed by observations of a CuAr lamp for wavelength
calibration and a quartz lamp for flat-fielding and slit location.
The net result
was that all of the targets received at least 90 minutes of exposure,
whilst the faintest targets, present in all three masks, were observed
for a total of 9 hours.

We used slits of width 0.5 arcsec and with lengths of 
8--10 arcsecs. The R831 grating was used with a long-pass OG515 filter
to block second order contamination. The resolving power was 4400 and
simultaneous sky subtraction of the spectra was possible. The spectra
covered $\sim 2000$\AA, with a central wavelength of 6200\AA --
7200\AA\ depending on slit location within the field of view.

The spectra were recorded on three $2048\times4068$ EEV chips leading to
two $\simeq 16$\AA\ gaps in the coverage. The CCD pixels were binned
$2\times2$ before readout, corresponding to $\sim 0.67$\AA\ per binned
pixel in the dispersion direction and 0.14 arcsec per binned pixel in
the spatial direction. Conditions were clear with seeing of 0.5--0.8
arcsec (FWHM measured from the spectra).

The data were reduced using version 1.8.1 of the GMOS data reduction
tasks running with version 2.12.2a of the Image Reduction and Analysis
Facility ({\sc iraf}). The data were bias subtracted, mosaiced and
flat-fielded. A two-dimensional wavelength calibration solution was
provided by the arc spectra and then the target spectra were
sky-subtracted and extracted using 2 arcsec apertures. The three
individual spectra for each target were combined using a rejection
scheme which removed obvious cosmic rays.  The instrumental
wavelength response was removed from the combined spectra using
observations of a white dwarf standard to provide a relative flux
calibration. The same calibration spectrum was used to construct a
telluric correction spectrum. A scaled version of this was divided into
the target spectra, tuned to minimise the RMS in regions dominated by
telluric features. The combined spectra were corrected to the
heliocentric reference frame and where multiple observations of a
target were obtained on more than one occasion through the same mask or
through different masks, these were tested for radial velocity
variations (see below) before combining into a single summed spectrum
for each target.

The SNR of each summed
spectrum was estimated empirically from the RMS deviations of straight
line fits to segments of ``pseudo-continuum'' close to the
\lii~6708\AA\ features (see below). As small unresolved spectral
features are expected to be part of these pseudo-continuum regions,
these SNR estimates, which range from $\sim 20$--30 in the faintest
targets to $>100$ in the brightest, should be lower limits. Examples of
the reduced spectra are shown in Fig.~\ref{specplot}. All the reduced
spectra are available in ``fits'' format from the ``Cluster'' Collaboration's home page
(see footnote 1).

\subsection{Analysis}
\label{analysis}
Each spectrum was analysed to yield a spectral type, equivalent widths
of the \lii~6708\AA\ and H$\alpha$ lines and a heliocentric RV.
Each of these analyses is described below. The results are given in
Table~\ref{specresults}.

\begin{table*}
\caption{
Results from the spectroscopic analyses. Columns list
the identifiers from Table~\ref{ccd_catalogue}, 
photometry, the signal-to-noise ratio of the summed spectra, the TiO
index, derived spectral type (on a  numerical scale where $-2=$K5,
$-1=$K7, $0=$M0, $1=$M1 etc.), the H$\alpha$ equivalent width (negative
$=$ absorption), the radial velocity, dispersion in the radial velocity
from multiple measurements and number of spectra/radial velocity
measurements, the lithium equivalent width (or 3-sigma upper limit) and its
uncertainty.}
\begin{tabular}{c@{\hspace*{2mm}}ccccccccccc}
\hline
\hline
CCD & ID & $I_C$ & $R_C-I_C$ & SNR & TiO & SpT & H$\alpha$ EW & RV &
$\sigma$RV & nRV & Li EW \\
&& \multicolumn{2}{c}{(mag)} &     &     &     & (\AA)
&\multicolumn{2}{c}{(km\,s$^{-1}$)} & & (\AA) \\
\hline
1.04& 827& 16.583& 0.920&  92&  1.16& -0.6&   2.0&  -7.2&  2.5& 3& $0.53\pm 0.02$\\
1.04& 829& 16.591& 0.875&  61&  1.11& -1.0&   0.2&  -9.4&     & 1& $0.31\pm 0.03$\\
1.04& 876& 16.508& 0.841& 133&  1.09& -1.2&   1.7&   5.9&  2.1& 3& $0.56\pm 0.01$\\
1.04&1018& 16.806& 0.896& 119&  1.01& -1.9&  -1.5&  -2.0&  2.0& 3& $<0.04$\\
1.04&1025& 16.878& 0.889& 218&  1.01& -2.0&  -2.8& -52.6&  4.2& 3& $<0.03$\\
1.04&1042& 16.729& 0.939&  64&  1.14& -0.8&   0.7&  -1.3&     & 1& $0.40\pm 0.03$\\
1.04&1056& 16.979& 1.056&  86&  1.23& -0.1&   2.4&  -6.8&  1.7& 2& $0.34\pm 0.02$\\
1.04&1269& 17.284& 1.146& 124&  1.32&  0.6&   1.5&  -7.1&  1.7& 3& $0.12\pm 0.01$\\
1.04&1291& 17.181& 1.234&  77&  1.46&  1.5&   5.0&  -3.0&  2.7& 3& $0.47\pm 0.02$\\
1.04&1540& 17.678& 1.554&  79&  1.80&  3.1&   6.1&  -1.7&  3.0& 2& $<0.06$\\
1.04&1545& 17.509& 1.322&  29&  1.51&  1.7&   2.9&  -5.1&     & 1& $<0.17$\\
1.04&1833& 17.357& 1.219& 153&  1.43&  1.3&   3.5&  -7.8&  1.9& 3& $0.15\pm 0.01$\\
1.04&1859& 17.863& 1.542&  91&  1.74&  2.9&   5.0&  -6.8&  4.0& 6& $<0.05$\\
1.04&1860& 17.853& 1.535&  63&  1.83&  3.2&   5.2&  -6.9&  1.2& 2& $<0.08$\\
1.04&1871& 18.015& 1.549&  35&  1.75&  2.9&   3.8&  -5.6&  0.6& 3& $<0.14$\\
1.04&1878& 17.712& 1.533& 100&  1.63&  2.4&   5.5&  -4.0&  4.4& 3& $<0.05$\\
1.04&2160& 18.132& 1.544&  71&  1.92&  3.6&  -0.3&  -9.1&     & 1& $<0.07$\\
1.04&2171& 18.191& 1.625&  93&  1.91&  3.5&   5.8&  -5.1&  2.6& 6& $<0.05$\\
1.04&2173& 18.334& 1.625&  81&  1.95&  3.6&   4.6&  -5.3&  8.4& 6& $<0.06$\\
1.04&2188& 18.179& 1.581&  37&  1.88&  3.4&   3.3&  -7.7&  0.6& 3& $<0.13$\\
1.04&2214& 18.213& 1.587&  50&  1.89&  3.4&   3.9&  -3.7&  2.3& 3& $<0.10$\\
1.04&2249& 18.270& 1.747&  41&  2.10&  4.1&   4.6&  -1.2&  1.1& 3& $<0.12$\\
1.04&2663& 18.412& 1.699&  46&  2.06&  4.0&   6.6&  -1.0&  1.7& 6& $<0.10$\\
1.04&2672& 18.658& 1.733&  50&  2.07&  4.0&   6.3&  -0.4&  1.9& 5& $<0.10$\\
1.04&2696& 18.584& 1.695&  24&  2.02&  3.9&   5.0&  -2.1&  1.1& 3& -- \\
1.04&2703& 18.490& 1.694&  51&  1.95&  3.6&   6.4&  -9.7&  2.1& 6& $<0.10$\\
1.04&3028& 19.060& 1.919&  40&  2.29&  4.5&   5.1&  -1.5&  2.4& 6& $<0.12$\\
1.04&3073& 18.785& 1.801&  45&  2.22&  4.4&   6.7&  -0.1&  2.3& 6& $<0.11$\\
1.04&3080& 18.795& 1.896&  34&  2.43&  4.8&   6.9&  -8.8&  5.0& 6& $0.69\pm 0.05$\\
1.04&3081& 18.711& 1.711&  33&  2.24&  4.4&   4.7&  -4.8&  3.1& 6& $0.17\pm 0.05$\\
1.04&3150& 18.843& 1.827&  52&  2.27&  4.5&   6.2&  -8.3&  2.2& 6& $<0.09$\\
1.04&3590& 19.107& 1.864&  31&  2.31&  4.6&   5.8&  -2.3&  3.8& 6& $<0.15$\\
1.04&3596& 19.185& 1.903&  29&  2.33&  4.6&   6.4&  -5.3&  0.5& 6& $0.62\pm 0.05$\\
1.04&3612& 19.152& 1.914&  33&  2.13&  4.2&   5.0&   4.5&  0.7& 6& $<0.15$\\
1.04&4165& 19.469& 1.992&  26&  2.28&  4.5&   4.1&   1.6&  6.6& 6& $<0.19$\\
\hline
\hline
\end{tabular}
\label{specresults}
\end{table*}

\subsubsection{Spectral Types}
\label{spt}

\begin{figure}
\includegraphics[width=80mm]{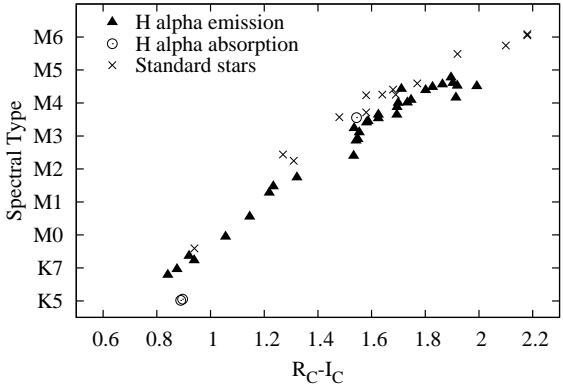}
\caption{Spectral types deduced from the TiO(7140\AA)
  index as a function of $R_C-I_C$. Three stars with H alpha absorption
  lines (and presumably non-members -- Section~\ref{member}) are indicated.
  Also shown are results from the spectra
  of standard stars where $R_C-I_C$ is available (see text).}
\label{sptri}
\end{figure}

\begin{table}
\caption{The relationship between TiO(7140\AA) index and spectral type
calibrated from standards in Montes et al. (1997) and Barrado y
Navascu\'es et al. (1999).}
\begin{tabular}{cc}
\hline
\hline
Index & Spectral Type \\
\hline
0.99 & K5 \\
1.13 & K7 \\
1.26 & M0 \\
1.40 & M1 \\
1.53 & M2 \\
1.74 & M3 \\
2.08 & M4 \\
2.61 & M5 \\
3.38 & M6 \\
\hline
\hline
\end{tabular}
\label{sptindex}
\end{table}

Spectral types were estimated from the strength of
the TiO(7140\AA) narrow band spectral index (e.g. Brice\~no et
al. 1998; Oliveira et
al. 2003). This index is temperature sensitive and
calibrated using the spectral types of well
known late-K and M-type field dwarfs taken from spectra in Montes et
al. (1997) and Barrado y Navascu\'es et al. (1999). We constructed a polynomial
relationship between spectral type and the TiO(7140\AA) index that was
used to estimate the spectral type of our targets, based on a
numerical scheme where M0--M6$=$0--6, K5\,$=-2$ and
K7\,$=-1$. Table~\ref{sptindex} gives the adopted relationship between
the TiO(7140\AA) index and spectral type. The scatter
around the polynomial indicates that these spectral types are good to
$\pm$ 0.3 subclasses for stars of type M1 and later, but about twice this
for earlier spectral types where the molecular bands are weak.

A plot of spectral type, from the TiO(7140\AA) index, versus
$R_C-I_C$ colour reveals a smooth relationship (see
Fig.~\ref{sptri}) with little scatter.
The most likely contaminants among our candidate members are foreground
M-type field dwarfs with similar spectral types but lower luminosities
or background K-giants. It is
possible that these could have different reddening that might make them
stand out in this diagram, but no objects exhibit a significant deviation. 
A comparison of the positions of some of the standard
stars on this plot ($R_C-I_C$ colours where available are from Leggett
1992) reveals an average redward offset of $\simeq 0.1$ mag in the $R_C-I_C$
values of our targets at a given spectral type. We
expect cluster members to have suffered a reddening $E(R_C-I_C)\simeq 0.14$ mag.
Whilst this comparison provides some evidence that the photometric 
calibration for these
red stars is reasonable, we have to temper this conclusion
with the possibility that the relationship between colour and spectral
type could alter for PMS stars with lower surface gravity.

\subsubsection{H-alpha measurements}

\begin{figure}
\includegraphics[width=80mm]{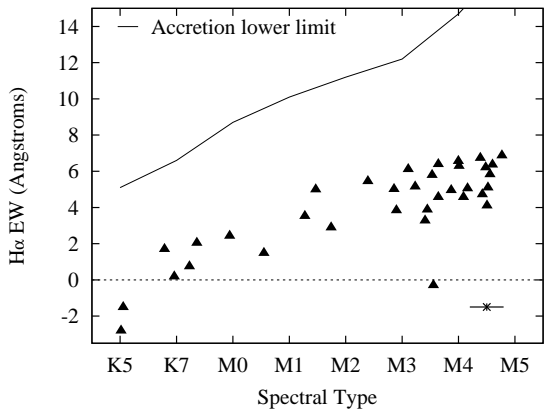}
\caption{Equivalent width (EW) of the H$\alpha$ line versus 
spectral types. A typical error bar is indicated. The majority of stars 
show H$\alpha$ emission commensurate with chromospheric activity from
young stars. None of the targets has H$\alpha$ emission characteristic
of accretion: the solid line shows the minimum EW expected from
accretion activity (defined by Barrado y Navascu\'es \& Mart\'in
2003). Three targets have H$\alpha$ absorption and are unlikely to be
cluster members (Section~\ref{member}).}
\label{haspt}
\end{figure}

H$\alpha$ equivalent widths (EWs) were
measured by direct integration 
above (or below) a pseudo-continuum.
The main uncertainty here is the definition of the pseudo-continuum as a
function of spectral type and probably results in uncertainties of order
0.2\AA, even for the bright targets. 

H$\alpha$ emission is ubiquitous from young stars. It arises either as
a consequence of chromospheric activity or is generated by accretion
activity in very young objects (e.g. Muzerolle, Calvet \& Hartmann
1998). It is unlikely that accretion persists in stars much beyond
10\,Myr (e.g. Jeffries et al. 2007; Fedele et al. 2010). The H$\alpha$
emission from accreting ``classical'' T-Tauri stars (CTTS) is
systematically stronger and broader than the weak line T-Tauri stars
(WTTS) where the emission is predominantly chromospheric.  

Fig.~\ref{haspt} shows the H$\alpha$ EW as a function of spectral
type for our targets.
None has H$\alpha$ emission strong enough to indicate accretion
according to criteria defined by White \&
Basri (2003) and Barrado y Navascu\'{e}s \& Mart\'{i}n (2003), but
the majority have emission characteristic of the chromospheric activity
expected for low-mass stars with an age of $\simeq 20$--$30$\,Myr
(e.g. Stauffer et al. 1997). Three
stars have H$\alpha$ absorption lines and are unlikely to be cluster
members (see Section~\ref{member}).

\subsubsection{Radial Velocities}

Our observations of each target were split into 1--6 epochs, depending
on in which masks the target featured. This gave the opportunity to
check for binarity, or at least binaries with orbital periods shorter than a
few months, by looking for RV variations.

Relative RVs were determined using the {\sc iraf} procedure {\sc fxcor}
to cross-correlate the first spectrum in a sequence of target exposures
with the rest, yielding between 0 and 5 RV difference measurements. All
spectra were heliocentrically corrected before correlation. For the
earlier type stars in our sample we found that the strongest
cross-correlation functions were obtained in the wavelength range
6000--6500\AA, though this range was truncated, at the blue end, at
longer wavelengths for some targets where the spectrum fell off the CCD
image. Similarly we found that the wavelength range 6600-7000\AA\ gave
the best correlations for the cooler targets. In practice, the division
between warmer and cooler stars was not made absolute and we took an
average of the two measurements for stars with spectral types between
M2.0 and M3.5. Statistical uncertainties in each RV should be of
order $\sim 100$/SNR km\,s$^{-1}$, but the dispersions for each target
are larger than this. A probable cause is small slit mis-centering
  errors of order a few hundredths of an arcsecond, exacerbated 
  by the good seeing compared to the
  slit width. In any case, the measured dispersions of a few km/s are a
better estimate of the true uncertainties.

No measurements of RV standards were taken as part of our observing
program, but heliocentric RVs were estimated by cross-correlating against
a synthetic spectral library in the same wavelength ranges discussed
above (generated from the Phoenix models by Brott \& Hauschildt
2005). write in a synthetic way
For the warmer stars we used the synthetic template of a solar
metallicity star at 4000\,K and with $\log g = 4.5$; for the cooler
stars we chose a synthetic template with solar metallicity, 3500\,K and
$\log g = 4.0$. Again, results were averaged in the overlap region.

Table~\ref{specresults} gives our estimate of the heliocentric RV (the
mean of results from each spectrum) and a
standard deviation where multiple relative RV measurements are available.
Given the likely uncertainties in each RV measurement, there is no
evidence, other than perhaps for target 1.04\_2173, that any of the
targets have RVs that
vary by more than a few km\,s$^{-1}$ on timescales of a month or less.

\subsubsection{Lithium measurements}
\label{li}

\begin{figure}
\includegraphics[width=80mm]{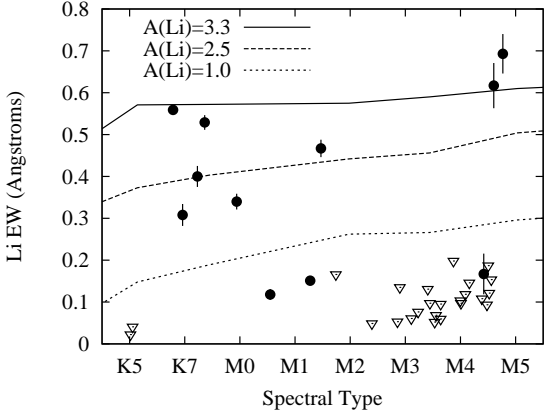}
\caption{Equivalent width (EW) of the Li\,{\sc i} 6708\AA\ feature versus 
spectral type. Triangles indicate 3-sigma upper limits. The loci
represent lines of constant Li abundance (A(Li) = 12 + $\log
N$(Li)/$\log N$(H)) calculated from the curves of growth described by
Jeffries et al. (2003) and the relationship between spectral type and
temperature presented by Kenyon \& Hartmann (1995).}
\label{liewplot}
\end{figure}

The \lii~6708\AA\ resonance feature should be strong in cool young
stars with undepleted Li -- an EW of 0.3\AA\ to
0.6\AA\ is predicted by curves of growth (e.g. Zapatero-Osorio
et al. 2002; Jeffries et al. 2003). Theory suggests (e.g. Baraffe et
al. 1998) that in a population with age $\sim
20$\,Myr, the Li EW should grow towards late K-type as the line
strengthens for a given abundance. Li-depletion in early- to mid-M
dwarfs should result in an undetectable Li line, and then below an
age-dependent luminosity, objects should have retained all their
original Li and the line EW should return sharply to $\sim 0.6$\AA.
Insets in Fig.~\ref{specplot} show the Li region in a
number of our targets.

Where it appeared, 
the EW of the \lii~6708\AA\ feature was estimated by direct integration 
below a pseudo-continuum derived from fitting small regions either side
of the Li line, excluding regions beyond 6712\AA\ which contain a
strong Ca line and which are noisy due to the subtraction of a strong
S\,{\sc ii} sky line. 
Uncertainties in the EW were estimated using
the formula $\Delta {\rm EW} = 1.6\sqrt{fp}/{\rm SNR}$ (Cayrel 1988),
where $f$ is the 
FWHM ($=1.5$\AA) of the unresolved line  and $p$ is the pixel size
($=0.67$\AA). In many cases
there was no obvious Li feature to measure, in which case a $3\sigma$
($= 3\, \Delta {\rm EW}$)
upper limit is quoted. In one case the Li feature fell in a gap between
the detectors and no EW could be measured. Li EWs are plotted as a
function of spectral type in Fig.~\ref{liewplot}. According to the
curves of growth described by Jeffries et al. (2003), the measured EWs
imply Li abundances from A(Li)$\simeq 3.3$ (where A(Li)
= 12 + $\log N$(Li)/$\log N$(H)), corresponding to the undepleted
meteoritic value (Anders \& Grevesse 1989), to A(Li)$\ll 1.0$.

\section{The Lithium Depletion Boundary}

\subsection{Cluster membership and sample contamination}
\label{member}

Determining the LDB requires a clean sample of genuine cluster members.
Several lines of evidence suggest that the vast majority of the objects
targeted for spectroscopy are members of NGC 1960. 

\subsubsection{Photometric selection}
An upper limit to the contamination of the
spectroscopic sample is estimated by comparing the spatial density of
objects inside the photometric selection box (in the $I_C$ versus
$R_C-I_C$ CMD) close to the cluster centre, where the targets were
selected, with the spatial density far from the cluster centre.  It is
an upper limit because we can not be sure that very low-mass cluster
members are confined only to the central region.  Sharma et al. (2006)
fitted a King model to brighter stars ($V<18$) in NGC~1960, finding a
core radius of only 3.2 arcminutes. If the lower mass stars follow the
same profile, then their spatial density should decrease by a
factor of 10 only $\sim 10$ arcminutes from the cluster centre. This
analysis used a ``background box'' of size 104 square
arcminutes about 10 arcminutes from the cluster centre (see
Fig.~\ref{gaiaplot}). Using the same criteria used to select Gemini
targets, there are 41 cluster candidates in this box, 27 with
$0.8<R_C-I_C\leq 1.4$ and 14 with $1.4<R_C-I_C<2.0$. This compares with
the GMOS field, with an effective area of 30 square arcminutes, that
contains 37 and 97 candidates in the same colour ranges, of which we
spectroscopically observed 11 and 24 respectively. If we assume the
background box contains only contaminants and that their density 
is constant across our survey, we expect 2.3 objects in
our spectroscopic sample with $0.8<R_C-I_C\leq 1.4$ to be non-members
and a further 1.0 non-member with $1.4<R_C-I_C<2.0$. 

\subsubsection{Lithium}
Eight of our targets have clear detections of the Li feature with
EW\,$>0.3$\AA. Comparisons with Li-depletion patterns in open clusters
of known age (e.g. see Fig.~10 of Jeffries et al. 2003 and references
therein) place empirical age constraints on these stars. Li EWs of
$>0.3$\AA\ are not seen in stars cooler than spectral type K5 in the
Pleiades or Alpha Per clusters with ages of 120\,Myr and 90\,Myr
respectively (excepting the very low luminosity M6$+$ stars
beyond the LDB, where Li remains unburned). Nor can strong Li lines be
seen in M dwarfs of the 35--55\,Myr open clusters NGC~2547, IC~2391 and
IC~2602 (again, excepting M4.5$+$ dwarfs beyond the LDB).  Thus objects
with EW[Li]\,$>0.3$\AA\ are probably younger than 100\,Myr, and younger
than 50\,Myr if they have spectral type M0--M5. These Li-rich objects
are very likely to be members of NGC 1960 as the chances of a
field star being younger than 100\,Myr is of order 1 per cent. However
a lack of Li cannot exclude candidates because we are
observing stars that are cool enough that even at an age of 20\,Myr we
might expect all their Li to have been depleted, especially at spectral
types M2--M4.

\subsubsection{Chromospheric emission}
H$\alpha$ measurements are a powerful way of excluding non-members.
For instance in the Pleiades, at an age of 120\,Myr,
all late K- and M-dwarf members show chromospheric H$\alpha$
emission (Stauffer et al. 1997). However, the H$\alpha$ magnetic
activity lifetimes of M-dwarfs range from a few hundred Myr in early
M-dwarfs to almost 5\,Gyr for M5 dwarfs
(West et al. 2008), so there is a significant probability that
contaminating foreground field M-dwarfs would still show H$\alpha$
emission.  Therefore, the 3 objects in our sample that
have H$\alpha$ absorption are either older foreground field dwarfs
or background giants, but the H$\alpha$ emission we see in all 
other targets is a necessary, but not sufficient, condition for
membership.

\subsubsection{Radial velocities} 
We expect cluster members to have similar RVs, 
with a dispersion of a few
km\,s$^{-1}$. Objects with RVs outside this range are either
non-members or possibly cluster members in binary systems.
Only one object (ID 1.04\_1025) has an RV clearly discrepant from the bulk of
objects. This target also has H$\alpha$ absorption, so is not a cluster
member in any case. None of the objects, except perhaps $1.04\_2173$, 
show evidence for any RV variability.

The unweighted mean heliocentric RV of the 8 objects with EW(Li)$>0.3$\AA
is $-5.1 \pm 1.5$\,km\,s$^{-1}$, with a standard deviation of
4.2\,km\,s$^{-1}$. If we take the whole sample, but exclude the 
three objects with H$\alpha$ 
absorption, the other 32 targets have an unweighted mean
heliocentric RV of $-4.0
\pm 0.7$\,km\,s$^{-1}$, a standard deviation of
3.8\,km\,s$^{-1}$, and all have RVs within 3 standard deviations of this mean.
It is therefore impossible to exclude further objects on the basis of
RV with any confidence and the RVs support the
idea that the majority of targets share a similar RV. 

In summary 3 targets (IDs 1.04\_1018, 1.04\_1025,
1.04\_2160) are excluded as non-members. The rest have H$\alpha$ emission, RVs,
spectral types and colours consistent with cluster
membership. Those with Li in their atmospheres
are probably younger than 100\,Myr and almost certain cluster
members. The level of sample contamination is as expected from the
photometric selection criteria.

\subsection{Locating the lithium depletion boundary}
\label{ldbloc}

\begin{figure}
\includegraphics[width=76mm]{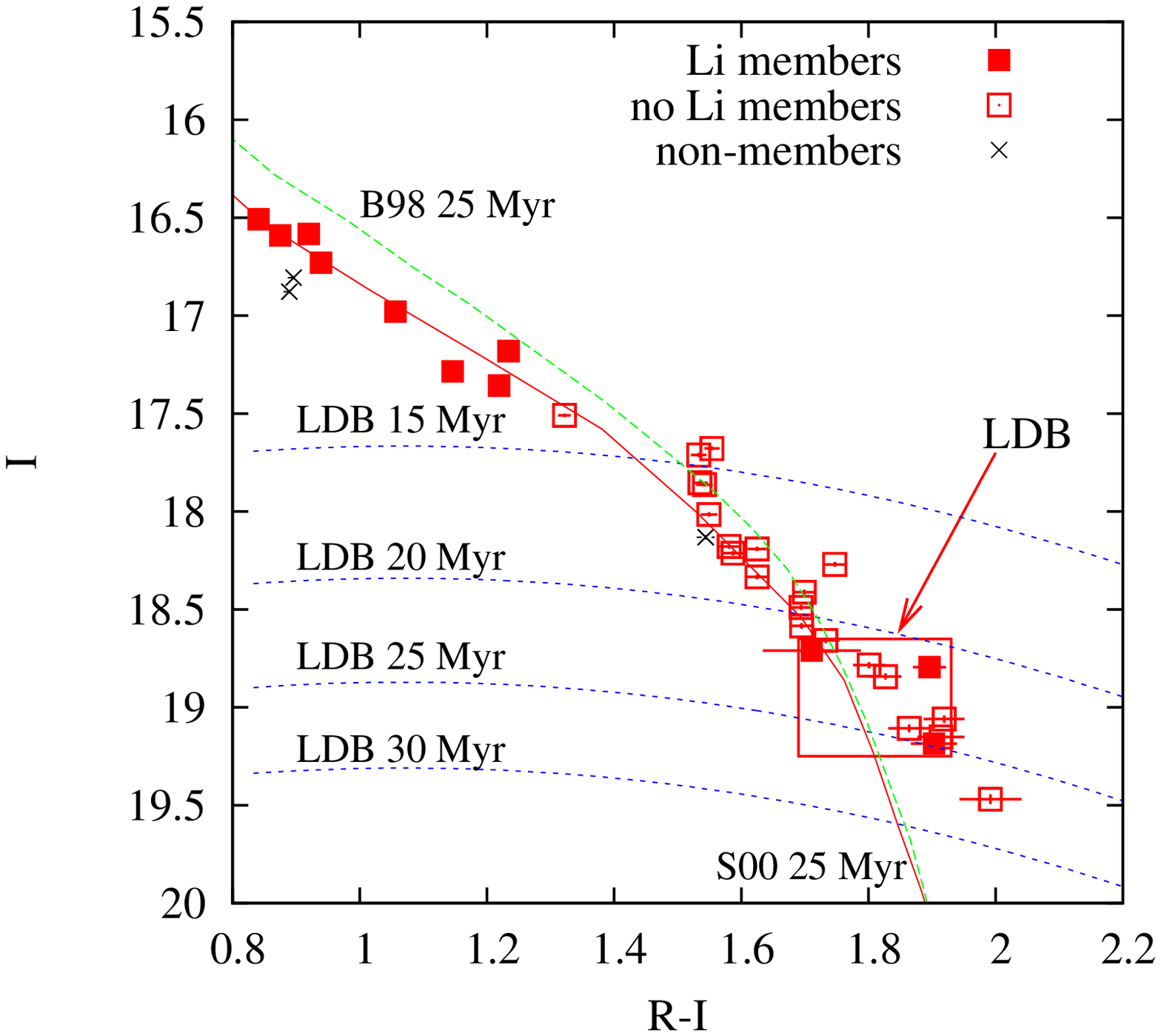}
\includegraphics[width=78mm]{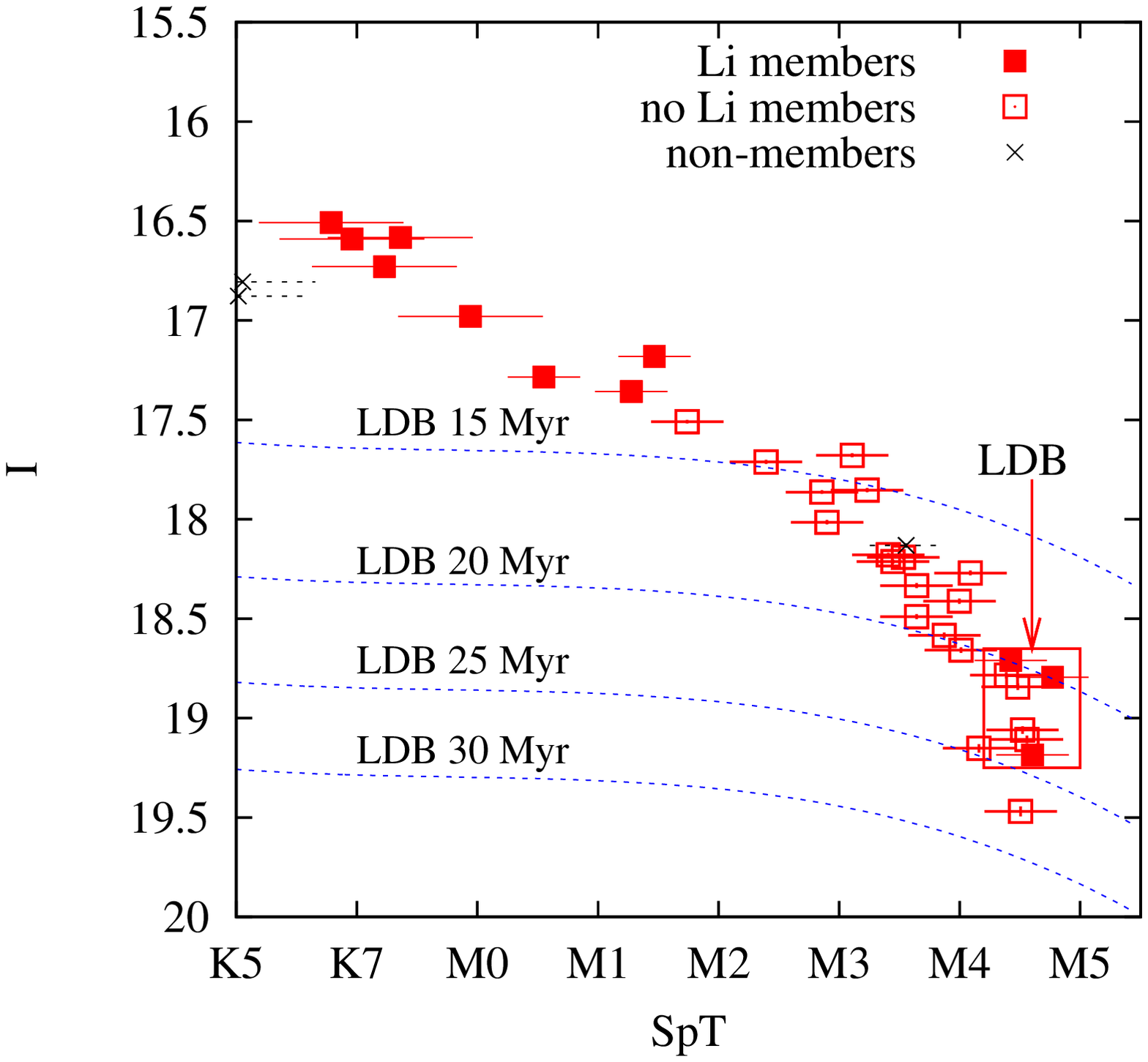}
\includegraphics[width=76mm]{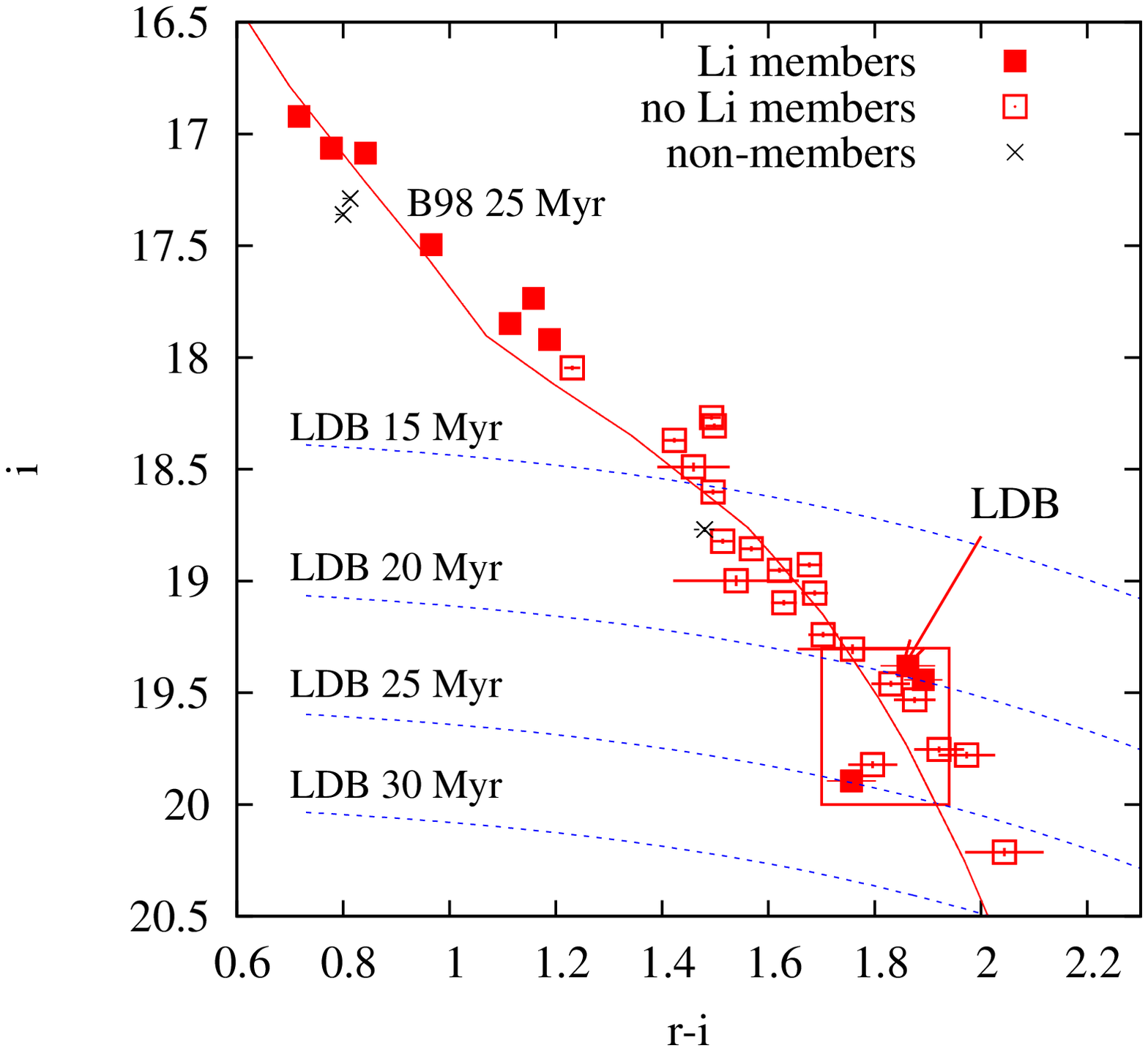}
\caption{Colour-magnitude and spectral type-magnitude diagrams
  indicating which objects have detected lithium and the
  likely location of the lithium depletion boundary. The dashed lines
  in each diagram are the predicted loci of 99 per cent Li
  depletion according to the models of Chabrier \& Baraffe (1997), using
  empirical bolometric corrections, reddening and extinction
  corresponding to $E(B-V)=0.20$ and a distance modulus of 10.33 mag 
  (see Section~\ref{ldbage}). The top panel also shows 25 Myr
  isochrones calculated (as described in Section~3) 
  from the interior models of Baraffe et al. (1998) and Siess et al. (2000). The
  bottom panel shows an equivalent isochrone calculated by Bell et al. (2013)
  from the Baraffe et al. models. 
}
\label{ldbplot}
\end{figure}

Fig. \ref{liewplot} shows that no stars have detected Li between
spectral types of M2 and M4 ($3580>T_{\rm eff}>3370$\,K -- Kenyon 
\& Hartmann 1995). According to theoretical models (e.g. D'Antona
\& Mazzitelli 1997; Chabrier \& Baraffe 1997; 
Siess et al. 2000), at cooler temperatures
the core temperatures of PMS stars remain too cool to burn Li, with
an abrupt transition occurring in Li abundance between entirely
depleted Li on the warm side of the boundary and undepleted Li on the
cool side. This does appear to be the case in our data (see
Fig.~\ref{liewplot}). The transition
occurs at a spectral type of $\simeq$M4.5, with the two coolest objects
(according to their spectral types, though not according to their
colours) showing undepleted Li levels.

In principle, the sharp transition in Fig.~\ref{liewplot} can be used
to estimate the cluster age. In practice, the $T_{\rm eff}$ or
spectral type of the LDB is {\it not} the best age indicator. As
explained in Jeffries (2006), there are significant uncertainties (of
order 150\,K) in converting a spectral type or colour into $T_{\rm
  eff}$, and different evolutionary models, using different atmosphere
prescriptions, differ by a similar amount in the $T_{\rm eff}$ 
predicted for the LDB at a given age. The relatively shallow relationship
  between $T_{\rm eff}$ at the LDB and age means
  that any small temperature uncertainty translates into a large
  age uncertainty. For example, an LDB at $3300\pm 150$\,K,
  corresponding to a spectral type of M4.5, leads to an 
  LDB age estimate of $31 \pm
  20$\,Myr via the models of Chabrier \& Baraffe (1997). As we show below, for NGC~1960 where the distance
  is reasonably well-determined, the LDB age is much more
  precisely estimated using the luminosity or absolute magnitude of the
  LDB.  Different evolutionary models also predict very similar LDB
luminosities at ages between 15\,Myr and 150\,Myr and relationships
between bolometric correction and colour are uncertain by $<0.1$\,mag,
which turns out to have a negligible effect on age estimates (Jeffries
\& Naylor 2001).

Figure~\ref{ldbplot} shows the CMDs and spectral-type versus magnitude
diagrams for our targets, indicating those with and without detected Li
and those that are non-members. In each diagram an estimate is
made of the colour (or spectral type) and magnitude range that marks the
LDB transition between stars that have depleted more than 99 per
cent of their initial Li, and those below that have retained Li.
There are difficulties in chosing this location. (i) It is
possible that some non-members (without Li) still lurk among the
low-mass stars, although this seems unlikely to be more than 1-2
objects given the discussion in Section~\ref{member}. (ii) Although
close binarity is unlikely in any of our targets, at the distance of
NGC~1960, even wide binaries with separations of several hundred au
would remain unresolved. The binary frequency among these low-mass
objects is expected to be of order 30 per cent. Binarity could increase
the apparent luminosity of a star at a fixed colour or spectral type
(for an equal mass binary) by 0.75 mag. Thus stars with Li could appear
up to 0.75 mag above the LDB. (iii) There are uncertainties in the
photometry and spectral types (shown in the plots) and young, low-mass stars can show
time-variable colours and magnitudes at levels of $\sim 0.1$ mag.

Of the three low-luminosity stars where Li has been detected, the star
1.04\_3080 shows some evidence of binarity, being $\sim 0.3$ mag brighter 
than the average object
at the same colour/spectral type in all three diagrams. 
The star 1.04\_3081 is the brightest of the three, but its EW(Li)
of $0.17\pm0.05$\AA\ suggests
it may already have  depleted $\sim 99$ per cent of its initial Li content.
We therefore define the LDB to lie in the boxes shown in
Fig.~\ref{ldbplot}, which allow a generous level of uncertainty.

\subsection{Ages from the LDB}

\label{ldbage}

The central points of the LDB boxes in Fig.~\ref{ldbplot} are
translated into ages as follows. The evolutionary models of Chabrier
\& Baraffe (1997) are interpolated to find a relationship between the
luminosity and age at which the initial Li content is 
depleted by 99 per cent. These luminosities are converted into absolute
magnitudes at any given colour or spectral type using empirical
bolometric corrections\footnote{We choose not to use the theoretical
  bolometric corrections from Baraffe et al. (1998) as these are known
  to poorly represent the optical colours of cool stars (e.g. Bell et
  al. 2012).}. 
For the $I_C$ versus $R_C - I_C$ diagram we use
a relationship between bolometric correction and colour obtained by
fitting a quadratic to data found in Leggett (1992) and Leggett et
al. (1996)
\begin{equation}
{\rm BC}_I = 0.174 + 0.882(R_C - I_C) - 0.4753(R_C - I_C)^2\, ,
\end{equation}
valid for $1.0 < R_C - I_C < 2.4$, with a scatter of 0.06 mag.
For the $I_C$ versus spectral type diagram we use a relationship
between BC$_{I}$ and spectral type derived from data given in Bessell
(1991)
\begin{eqnarray}
{\rm BC}_I & = & 0.595 -0.0108\,{\rm SpT} -0.00222\,{\rm SpT}^2 \nonumber
\\
           &   & -0.00342\,{\rm SpT}^3 ,
\end{eqnarray}
where SpT is the numerical spectral type given in
Table~\ref{specresults} and the relation is calibrated between types K7
and M6 with a scatter of 0.02 mag.
For the $i$
versus $r-i$ diagram we use a polynomial fit to gravity- and
temperature-dependent empirical bolometric
corrections calculated by Bell et al. (2012, 2013):
\begin{eqnarray}
{\rm BC}_i & = & -0.068 - 0.09797(r-i) +0.02586(r-i)^2 \nonumber \\
           &   &  -0.064736(r-i)^3\, ,
\end{eqnarray}
which is valid for $0.8 < r-i <2.5$ with a scatter of 0.04 mag.

The relationship between luminosity at the LDB and age combined with
the bolometric corrections, define constant luminosity isochrones in
the colour-magnitude or spectral type-magnitude diagrams. These
isochrones are compared with the observed data in Fig.~\ref{ldbplot} by
making the appropriate corrections for distance modulus (assumed to be
$10.33$) and reddening (either $E(R_C-I_C)=0.143$ or $E(r-i)=0.13$) and
extinction (either $A_I = 0.37$ or $A_i = 0.38$). These isochrones can
be interpolated to obtain the LDB age corresponding to any location in
the diagrams. This way of displaying the data and models has the
advantage of explicitly showing the influence of any particular choice
of LDB location or uncertainties in distance and reddening on the
derived age. 

Observational uncertainties in the LDB age are derived by
perturbing the LDB location by the uncertainties implied by the boxes
in Fig.~\ref{ldbplot}, by uncertainties in the distance modulus,
reddening and extinction and we also conservatively 
assume that the magnitudes and colours suffer from systematic uncertainties of
0.1 mag at these red colours and that the spectral type scale may have
systematic uncertainties of half a subclass. All these perturbations
are combined in quadrature to give 
uncertainties in the bolometric magnitude at the LDB and
consequent uncertainties in the LDB age.
This total uncertainty is
dominated by the size of the boxes in Fig.~\ref{ldbplot}. The
uncertainties due to distance, extinction, reddening and photometric
calibration are small in comparison. The relevant data and results
are presented in Table~\ref{ldbresults}.

Some idea of additional systematic errors can be gained from comparing the
bolometric magnitudes and ages deduced from the three separate
diagrams. There are differences 
of order 0.15 mag and 1.5\,Myr respectively, which are
smaller than the observational uncertainties. Additional model
dependencies are checked by: (i) Calculating the LDB age assuming that the
LDB location refers to the point at which Li is depleted by 90 or 99.9
per cent rather
than 99 per cent. This conservatively allows for an order of magnitude
uncertainty in the Li abundance predicted from a Li EW, but only changes 
the ages by $\pm 1.5$\,Myr, due to the rapid depletion
of Li once Li-burning is initiated. (ii) Calculating the
LDB ages using the models of D'Antona \& Mazzitelli (1997), 
Siess et al. (2000, with metallicity 0.02 or 0.01)
and Burke et al. (2004). These results are also given in
Table~\ref{ldbresults} for each of the three diagrams in
Fig.~\ref{ldbplot}. The use of alternative models changes the
derived age by $\pm 2$\,Myr, illustrating how insensitive the LDB age
is to choice of atmosphere, convection treatment or even factors of two
in metallicity.

Considering all the results, we give a final estimate for the LDB age of 
NGC 1960 as $22 \pm 3.5$\,Myr, where the observational uncertainty is
primarily associated with locating the LDB in sparse data. There is
then a further $\pm 2$\,Myr associated
with choice of evolutionary model and bolometric corrections, leading
to a final result of $22 \pm 4$\,Myr. It is
important to separate out these two uncertainty contributions, since 
the former could be almost eliminated by locating the LDB with more
precision.

\begin{table*}
\caption{The locations of the lithium depletion boundary in
  colour-magnitiude or spectral type-magnitude diagrams.  These
  locations are translated into a bolometric magnitude using an
  intrinsic distance modulus of 10.33 along with extinctions, reddening
  and bolometric corrections as described in
  Section~\ref{ldbage}. These bolometric magnitudes imply masses and LDB
  ages (from the models of Chabrier \& Baraffe 1997) as shown and LDB
  ages are also calculated for
  for a variety of other evolutionary models.}
\begin{tabular}{lccc}
\hline
\hline
 & $I_C$ vs $R_C - I_C$ & $I_C$ vs SpT & $i$ vs $r-i$ \\
\hline
LDB location  & $I_C=18.95 \pm 0.30$ & $I_C=18.95 \pm 0.30$ &
$i=19.65\pm 0.35$ \\
              & $R_C-I_C = 1.81 \pm 0.12$ & SpT $=$ M4.6$\pm 0.4$ &
$r-i=1.82\pm0.12$ \\

\hline

$M_{\rm bol}$ & $8.57\pm 0.33$ & $8.41 \pm 0.35$& $8.47 \pm 0.37$ \\
Mass ($M_{\odot}$) & $0.23\pm 0.04$ & $0.25\pm 0.04$ &$0.24\pm 0.04$ \\
\hline

Ages (Myr) &&&\\

Chabrier \& Baraffe 1997 & $23.2^{+3.5}_{-3.1}$ &
$21.6^{+3.5}_{-3.0}$& $22.2^{+3.8}_{-3.3}$ \\
D'Antona \& Mazzitelli 1997 & $19.8^{+3.8}_{-3.1}$ &
$18.3^{+3.6}_{3.0}$ & $18.7^{+4.1}_{-3.3}$ \\
Siess et al. 2000 (Z=0.02) & $23.1^{+3.8}_{-3.4}$ &
$21.5^{+3.8}_{-3.5}$ & $22.0^{+4.1}_{-3.8}$ \\
Siess et al. 2000 (Z=0.01) & $21.7^{+3.8}_{-3.3}$ &
$20.1^{+3.7}_{-3.2}$ & $20.6^{+4.1}_{-3.5}$ \\
Burke et al. 2004 & $21.3^{+3.7}_{-3.1}$ &
$19.8^{+3.5}_{-2.0}$ & $20.3^{+4.0}_{-3.2}$ \\
\hline
\hline
\end{tabular}
\label{ldbresults}
\end{table*}

\section{Discussion}

\subsection{The sharpness of the LDB}
NGC~1960 is the eighth cluster with an LDB age and also the
youngest. The data in Fig.~\ref{ldbplot} allow a reasonable estimate of
the LDB location, but in each case there is at least one star without
Li fainter than the adopted LDB and other Li-poor stars that share
similar locations to the two Li-rich (approximately undepleted)
low-mass stars.  This might be explained by a
combination of binarity, photometric uncertainties and contamination by
Li-poor non-members (see Section~\ref{ldbloc}).  However,
Fig.~\ref{ldbplot} shows that it would only take an age spread of $\sim
5$\,Myr within the NGC~1960 cluster to effectively blur the LDB
location and lead to mixing between the Li-poor and Li-rich
populations. Age spreads of this size are controversial, but may explain the
Hertzsprung-Russell diagrams of very young clusters (e.g. Palla \&
Stahler 2000, but see counter arguments in Hartmann 2001) and the
spread of Li depletion amongst low-mass stars ($M\simeq 0.1$ --
$0.3\,M_{\odot}$) in some star forming regions (Palla et al. 2007;
Sacco et al. 2007).  The LDB of NGC~2547, which is better defined than
that of NGC~1960 by a larger sample of Li-rich and Li-poor members,
also shows some evidence for this blurring (Jeffries \& Oliveira
2005). However, because NGC 2547 is older ($35\pm 3$\,Myr) than
NGC~1960, the LDB isochrones are closer together and the effects of any
genuine age spread are diminished with respect to those mimicked by binarity,
photometry errors and variability. NGC~1960 has more potential for
exploring the sharpness of the LDB in detail. Strong constraints on any
possible age spread might be found from measuring Li depletion in the many
tens of uninvestigated candidates in and around the LDB boxes in Fig.~\ref{ldbplot}.

\subsection{A robust age determination}
The richness of NGC~1960 has allowed (see
Section~2) statistically precise age estimates from fitting
isochrones to the upper main sequence and low-mass PMS. Cluster ages derived
from high-mass stars are influenced by physical factors such as the
amount of convective core overshoot, rotational mixing and mass loss
that are included in the evolutionary models (e.g. Maeder \& Meynet
1989; Schaller et al. 1992; Meynet \& Maeder 2000). Ages derived from
low-mass PMS models are affected by choices of
convection treatment, the equation of state and atmospheres (Siess
et al. 2000; Baraffe et al. 2002).  Both techniques are affected by a
choice of chemical composition, the way in which theoretical
luminosities and temperatures are transformed to compare
with observational data, via theoretical or empirical bolometric
corrections (or vice-versa).  There is also some role played by the
treatment of any binary population and the way in which
models are fitted to data (e.g. Naylor \& Jeffries 2006; von Hippel et
al. 2006).

It has been argued, given the long list of uncertainties above, that
LDB ages are more accurate than both upper main sequence and PMS
isochronal ages (Jeffries \& Naylor 2001; Burke et al. 2004), because
they circumvent or are much less sensitive to several observational
uncertainties, rely on physics that is considerably better
understood and are insensitive to choice of model or composition (see
Table~7). LDB ages are not entirely independent from
  ages determined by high- and low-mass isochronal fits, because they
  also require an adopted distance and reddening. However the
  LDB age estimated here is extremely insensitive to these
  parameters. An increase in distance modulus of 0.1 mag (twice its
  estimated uncertainty) would only decrease the LDB age by 1\,Myr and
  this insensitivity is qualitatively similar for all the LDB ages
  reported in the literature. LDB ages therefore offer the possibility
of calibrating out uncertainties in other methods and perhaps even
understanding what physical ingredients are responsible for any
discrepancies. In this way they could play a similar role for young
clusters ($<200$\,Myr) that white dwarf cooling chronometry is playing
in older clusters (De Gennaro et al. 2009; Jeffery et al. 2011).

\subsection{Concordance with other age estimates}
The LDB age of $22 \pm 4$\,Myr is consistent with previous estimates of
the cluster age based on isochronal fits to the upper main sequence
(Sanner et al. 2000; Sharma et al. 2006 -- see Section 2). Most
recently, Bell et al. (2013) estimated an upper main sequence age of
$26.3^{+3.2}_{-5.2}$\, Myr, using the non-rotating models of Schaller
et al. (1992) and Lejeune \& Schaerer (2001), which incorporate
convective overshooting of 0.2 pressure scale heights for
$M>1.5M_{\odot}$. Some authors (e.g. Stauffer et al. 1999; Cargile et
al. 2010) have argued that agreement between LDB ages and main sequence
turn-off ages in older clusters ($\sim 100$\,Myr) like the Pleiades and
Blanco 1, requires overshooting since without it turn-off
ages would be 30--40 per cent younger than LDB ages. The age derived
for NGC~1960 by Bell et al. (2013) comes from the rate of progression
from the ZAMS to the terminal age main sequence (TAMS) rather than the
turn-off. The main effect of convective overshoot is to displace the
the ZAMS and TAMS redward (or to higher luminosities) in the CMD,
broaden the gap between ZAMS and TAMS, whilst leaving the shape of the
isochrones for main sequence stars almost unchanged (see Maeder \&
Meynet 1989). In the luminosity range fitted by Bell et al. (2013),
which is below the main sequence turn-off, a model with no overshoot
would yield a distance modulus greater by $\sim 0.2$ mag, but an
unchanged age. The altered distance would lead to a $\sim 2$\,Myr
younger LDB age. Hence models featuring no overshooting would result in
a mild disagreement between the LDB age and upper main sequence age.

Models with no convective overshoot are unlikely unless another
parameter, such as rotation, increases the width of the predicted main
sequence between ZAMS and TAMS to match that observed in field star
samples.  Meynet \& Maeder (2000), and more recently Ekstr\"om et
al. (2012), show that rotation broadens the main sequence
and extends main sequence lifetimes in a similar way to
overshooting. The rotating Geneva models of Ekstr\"om et al. (2012),
with rotation rates about 40 per cent of break up, but which still
incorporate 0.1 pressure scale heights of overshoot, have a ZAMS
fainter by 0.08 mag compared with the Lejeune \& Schaerer
(2001) models employed by Bell at al (2013).  After adjusting the
distance modulus for this small difference, the upper main sequence age
would be unchanged, the PMS age (see below) increases by about 3\,Myr
and the LDB age would increase by just 1\,Myr.  Hence at the level of
precision achieved, the LDB age of NGC~1960 is consistent with models
that incorporate a moderate amount of overshoot or rotation (or a bit
of both) and isolating these effects using LDB ages is likely to be
difficult.

Bell et al. (2013) also used low-mass (0.7--1.5\,$M_{\odot}$)
members of NGC 1960, selected on the basis of their radial velocities 
and the presence of lithium to fit PMS isochrones from Baraffe et al. (1998, the set
with a mixing length of 1.9 pressure scale heights), D'Antona \&
Mazzitelli (1997) and Dotter et al. (2008) in the $g$ versus $g-i$ CMD.
The ages determined were 19.0--20.9\,Myr for the Baraffe et al. and Dotter et
al. models and 17.4--19.1\,Myr for the D'Antona \& Mazzitelli
models. These ages are in good agreement with each other and 
the LDB age, despite considerable differences in the physics they
incorporate. The slightly lower age for the D'Antona \& Mazzitelli
isochrone mirrors the lower LDB age based on
those models (see Table~\ref{ldbresults}). The targets in this paper
extend to much lower masses than those considered by Bell et
al. (2013) and the photometric calibrations and bolometric corrections 
are more uncertain (we allowed an additional 0.1
mag uncertainty in the LDB colour and magnitude). 
Nevertheless, appropriately reddened 25\,Myr isochrones
adopted from the interior models of Baraffe et al. (1998) and Siess et
al. (2000), with colour-$T_{\rm eff}$ calibrations tuned to match the
Pleiades (see Section~3) give a reasonable match to the run of cluster
members in the $I_C$ vs $R_C-I_C$ CMD (see Fig.~\ref{ldbplot}). 
Similarly, a 25\,Myr isochrone
calculated using the Baraffe et al. (1998) models and semi-empirical
bolometric corrections from Bell et al. (2013) is a good match to cluster
members in the $i$ versus $r-i$ CMD.

\section{Summary}

NGC~1960 is a rich northern hemisphere cluster, where ages have been
previously determined by fitting isochrones to the high- and low-mass
populations.  In this paper we have presented a photometric survey that
has been used to select a sample of very low-mass candidate cluster
members and these candidates have been spectroscopically examined to
establish the luminosity at which lithium remains unburned in their
atmospheres. By examining a variety of membership indicators, it has
been established that there is little contamination in the sample and
the ``lithium depletion boundary'' (LDB) has been used to establish an
age of $22 \pm 4$\,Myr for NGC~1960, where most of the uncertainty is
associated with locating the LDB in colour-magnitude (or spectral
type-magnitude) diagrams.  The uncertainty associated with choice of
low-mass evolutionary model and empirical bolometric corrections is
limited to just $\pm 2$\,Myr.

The LDB age for NGC~1960 is in good agreement with recent, more
model-dependent, age determinations
from its upper main sequence and low-mass PMS populations.  This
overall agreement does not in isolation offer strong constraints
on the uncertain physical ingredients of the high- and low-mass
stellar models, although high-mass models without any convective overshoot
or rotation are not favoured. 
Nevertheless, this is the first demonstration of
concordance between all three of these techniques, offering some
encouragement that absolute cluster ages at $\sim 20$\,Myr can be
determined reliably from any of these methods.

\section*{Acknowledgements}
Based on observations made with the Isaac Newton Telescope operated
on the island of La Palma by the Isaac Newton Group in the Spanish
Observatorio del Roque de los Muchachos of the Instituto de
Astrofisica de Canarias. 

Based on observations obtained at the Gemini Observatory, which is
operated by the Association of Universities for Research in Astronomy,
Inc., under a cooperative agreement with the NSF on behalf of the
Gemini partnership: the National Science Foundation (United States),
the Science and Technology Facilities Council (United Kingdom), the
National Research Council (Canada), CONICYT (Chile), the Australian
Research Council (Australia), Minist\'{e}rio da Ci\^{e}ncia, Tecnologia
e Inova\c{c}\~{a}o (Brazil) and Ministerio de Ciencia, Tecnolog\'{i}a e
Innovaci\'{o}n Productiva (Argentina)

CPB acknowledges receipt of a Science and
Technology Facilities Council postgraduate studentship. SPL is
supported by a RCUK fellowship.

\nocite{maeder89}
\nocite{ekstrom12}
\nocite{dantona97}
\nocite{chabrier97}
\nocite{leggett92}
\nocite{leggett96}
\nocite{baraffe98}
\nocite{cutri03}
\nocite{bildsten97}
\nocite{ushomirsky98}
\nocite{brott05}
\nocite{west08}
\nocite{bell12}
\nocite{bell13}
\nocite{degennaro09}
\nocite{jeffery11}
\nocite{manzi08}
\nocite{jeffriescargese01}
\nocite{burke04}
\nocite{jeffriesn216907}
\nocite{stauffer98}
\nocite{muzerolle98}
\nocite{fedele10}
\nocite{stauffer99}
\nocite{white03}
\nocite{barrado03}
\nocite{barradoldb04}
\nocite{zapatero02}
\nocite{cargile10ldb}
\nocite{jeffries05}
\nocite{jeffries04}
\nocite{dobbie10}
\nocite{soderblom10}
\nocite{barkhatova85}
\nocite{johnson53}
\nocite{landolt92}
\nocite{sanner00}
\nocite{maeder76}
\nocite{sharma06}
\nocite{mayne08}
\nocite{schaller92}
\nocite{baraffe02}
\nocite{jeffries03}
\nocite{jeffries09ic4665}
\nocite{stetson87}
\nocite{kenyon95}
\nocite{siess00}
\nocite{palla00}
\nocite{naylor98}
\nocite{naylor02}
\nocite{burningham03}
\nocite{cayrel88}
\nocite{anders89}
\nocite{stauffer97ic23912602}
\nocite{jeffries06lireview}
\nocite{bessell91}
\nocite{hartmann01}
\nocite{palla07}
\nocite{sacco07}
\nocite{meynet00}
\nocite{naylor06}
\nocite{vonhippel06}
\nocite{lejeune01}
\nocite{dotter08}
\nocite{taylor86}
\nocite{oliveira03}
\nocite{montes97}
\nocite{barrado99}
\nocite{briceno98}

\label{lastpage}
\bibliographystyle{mn2e} 
\bibliography{iau_journals,master}

\end{document}

%% file: latex_macros.tex
\newcommand{\rmsub}[2]{#1_{\rm #2}} 
\newcommand{\alfven}{Alfv\'en}
%
%
\newcommand{\aat}{\mbox{\em AAT}}
\newcommand{\eso}{\mbox{\em ESO}}
\newcommand{\iue}{\mbox{\em IUE}}
\newcommand{\exosat}{\mbox{\em EXOSAT}}
\newcommand{\einstein}{\mbox{\em Einstein}}
\newcommand{\ginga}{\mbox{\em GINGA}}
\newcommand{\rosat}{\mbox{\em ROSAT}}
\newcommand{\caspec}{\mbox{\em CASPEC}}
\newcommand{\ucles}{\mbox{\em UCLES}}
\newcommand{\starlink}{\mbox{\em Starlink}}
\newcommand{\etal}{\mbox{\em et\ al.\ }}
\newcommand{\dex}[1]{\hbox{$\times\hbox{10}^{#1}$}}
\newcommand{\eex}[1]{\hbox{$\hbox{10}^{#1}$}}
\newcommand{\gpar}{\mbox{$g_{\parallel}$}}
\newcommand{\vsi}{\mbox{$v_e\,\sin\,i$}}
\newcommand{\vsini}{\mbox{$v_e\,\sin\,i$}}
\newcommand{\pattcit}[1]{\hbox{$^{(#1)}$}}
\newcommand{\pattcite}[1]{\hbox{$^{(#1)}$}}
%
%
\newcommand{\ha}{H$\alpha$}
\newcommand{\hb}{H$\beta$}
\newcommand{\hgam}{H$\gamma$}
\newcommand{\hdel}{H$\delta$}
\newcommand{\heps}{H$\epsilon$}
\newcommand{\lya}{\hbox{$\hbox{Ly}\alpha$}}
\newcommand{\naid}{\mbox{Na{\footnotesize I} {\sl D}}}
\newcommand{\caii}{Ca\,{\footnotesize II}}
\newcommand{\caiih}{Ca\,{\footnotesize II}~H}
\newcommand{\caiik}{Ca\,{\footnotesize II}~K}
\newcommand{\caiihk}{Ca\,{\footnotesize II}~H \&~K}
\newcommand{\mgii}{Mg\,{\footnotesize II}}
\newcommand{\mgiih}{\mbox{Mg{\footnotesize II} {\sl h}}}
\newcommand{\mgiik}{\mbox{Mg{\footnotesize II} {\sl k}}}
\newcommand{\mgiihk}{\mbox{Mg{\footnotesize II} {\sl h} \&\ {\sl k}}}
\newcommand{\lii}{Li\,{\footnotesize I}}
\newcommand{\fei}{Fe\,{\footnotesize I}}
\newcommand{\baii}{Ba\,{\footnotesize II}}
\newcommand{\ki}{K\,{\footnotesize I}}
\newcommand{\cai}{Ca\,{\footnotesize I}}
%
%
\newcommand{\ang}{\,\mbox{\AA}}
\newcommand{\Ang}{\ang}
\newcommand{\angstrom}{\ang}
\newcommand{\angstroms}{\ang}
\newcommand{\Angstrom}{\ang}
\newcommand{\Angstroms}{\ang}
\newcommand{\km}{\,km}
\newcommand{\Mpc}{\,\mbox{Mpc}}
\newcommand{\kpc}{\,\mbox{kpc}}
\newcommand{\kms}{\,km\,s$^{-1}$}
\newcommand{\ergs}{\,\mbox{$\mbox{erg}\,\mbox{s}^{-1}$}}
\newcommand{\ergsqcmsecang}{\,erg\,cm$^{-2}$\,s$^{-1}$\,\AA$^{-1}$}
\newcommand{\ergsqcm}{\,erg\,cm$^{-2}$}
\newcommand{\ergsqcmsec}{\,erg\,cm$^{-2}$\,s$^{-1}$}
\newcommand{\sqcm}{\,\mbox{$\mbox{cm}^{2}$}}
\newcommand{\cucm}{\,\mbox{$\mbox{cm}^{3}$}}
\newcommand{\persqcm}{\,\mbox{$\mbox{cm}^{-2}$}}
\newcommand{\gpersqcm}{\,\mbox{g}\persqcm}
\newcommand{\percc}{\,\mbox{$\mbox{cm}^{-3}$}}
\newcommand{\kev}{\,\mbox{keV}}
\newcommand{\kelvin}{\,K}
\newcommand{\kgmcube}{\,\mbox{$\mbox{kg}\,\mbox{m}^{-3}$}}
\newcommand{\dynsqcm}{\,\mbox{$\mbox{dyn}\,\mbox{cm}^{-2}$}}
\newcommand{\degrees}{\mbox{$^\circ$}}
\newcommand{\rstar}{\,\mbox{$\mbox{R}_*$}}
\newcommand{\mstar}{\,\mbox{$\mbox{M}_*$}}
\newcommand{\lstar}{\,\mbox{$\mbox{L}_*$}}
\newcommand{\vstar}{\,\mbox{$\mbox{V}_*$}}
\newcommand{\msun}{\,\mbox{$\mbox{M}_{\odot}$}}
\newcommand{\rsun}{\,\mbox{$\mbox{R}_{\odot}$}}
\newcommand{\lsun}{\,\mbox{$\mbox{L}_{\odot}$}}
\newcommand{\lx}{\,\mbox{$L_{\rm x}$}}
%
%
\newcommand{\reference}[5]{\noindent #1, #2. {\sl #3\/}, {\bf #4,} \,\mbox{#5}} 
\newcommand{\refnum}[4]{\noindent #1, #2. {\sl #3\/}, {\bf #4}}
\newcommand{\refbook}[3]{\noindent #1, #2. {\sl #3\/}}
\newcommand{\refpress}[3]{\noindent #1, #2. {\sl #3\/}, in press}
\newcommand{\refsub}[3]{\noindent #1, #2. {\sl #3\/}, submitted}
\newcommand{\refprep}[2]{\noindent #1, #2. In preparation}
%
%
\newcommand{\aanda}  {Astr.\ Astro\-phys.\nolinebreak\ }
\newcommand{\aasupp} {Astr.\ Astro\-phys.\nolinebreak\ Suppl.\nolinebreak\ }
\newcommand{\aj}     {Astron.\nolinebreak\ J.\nolinebreak\ }
\newcommand{\annrev} {Ann.\ Rev.\ Astr.\ Astro\-phys.\nolinebreak\ }
\newcommand{\acta}   {Acta Astron.\nolinebreak\ }
\newcommand{\apj}    {Astro\-phys.\nolinebreak\ J.\nolinebreak\ }
\newcommand{\apjs}   {Astro\-phys.\nolinebreak\ J.\ Suppl.\nolinebreak\ }
\newcommand{\apjsupp}{\apjs}
\newcommand{\aplett} {Astro\-phys.\nolinebreak\ Lett.\nolinebreak\ }
\newcommand{\gafd}   {Geo\-phys.~Astro\-phys.\ Fluid Dyn.\nolinebreak\ }
\newcommand{\ibvs}   {Inf.\ Bull.\ var.\ Stars\nolinebreak\ }
\newcommand{\jgr}    {J.\ Geo\-phys.~Res.\nolinebreak\ } 
\newcommand{\jpp}    {J.\ Plasma Phys.\nolinebreak\ }
\newcommand{\mn}     {Mon.\ Not.\ R.\ astr.\nolinebreak\ Soc.\nolinebreak\ }
\newcommand{\pf}     {Phys.\nolinebreak\ Fluids\nolinebreak\ }
\newcommand{\pasp}   {Publ.\ astr.\ Soc.\ Pacif.\nolinebreak\ }
\newcommand{\sovast} {Soviet astr.\nolinebreak\ }
\newcommand{\procasa}{Proc.\ Astr.\ Soc.\ Australia\nolinebreak\ }
\newcommand{\solp}   {Solar Phys.\nolinebreak\ }
\newcommand\mnras{MNRAS}
\newcommand\aap  {A\&A}
\newcommand\aaps  {A\&AS}

%
%
\newcommand{\der}[1]{{\rm d}#1}
\newcommand{\deriv}[2]{\frac{{\rm d}#1}
                       {{\rm d}#2}} 
\newcommand{\sderiv}[2]{\frac{{\rm d}^2#1}\over
                       {{\rm d}#2^2}} 
\newcommand{\pderiv}[2]{\frac{\partial#1}
                       {\partial#2}} 
\newcommand{\spderiv}[2]{\frac{\partial^2#1}
                        {\partial#2^2}} 
\newcommand{\half}{\mbox{$\frac{1}{2}$}}
\newcommand{\twiddles}{\mbox{$\sim $}}
\newcommand{\varomega}{\varpi}
\newcommand{\twid}{\mbox{$\sim $}}
\newcommand{\bvec}[1]{\mbox{\boldmath ${#1}$}}
\newcommand{\be}{\begin{equation}}
\newcommand{\ee}{\end{equation}}
\newcommand{\bd}{\begin{displaymath}}
\newcommand{\ed}{\end{displaymath}}
\newcommand{\del}{{\bf \nabla}}